\documentstyle[preprint,prd,eqsecnum,aps,epsf,axodraw]{revtex}

\tightenlines 
\begin{document}

    \newcommand{\DSC}{D\hspace{-0.25cm}\slash_{\bot}}
    \newcommand{\DSP}{D\hspace{-0.25cm}\slash_{\|}}   
    \newcommand{\DS}{D\hspace{-0.25cm}\slash}   
    \newcommand{\DC}{D_{\bot}}
    \newcommand{\DSCX}{D\hspace{-0.20cm}\slash_{\bot}}
    \newcommand{\DSPX}{D\hspace{-0.20cm}\slash_{\|}}  
    \newcommand{\DP}{D_{\|}}
    \newcommand{\QV}{Q_v^{+}}
    \newcommand{\QVB}{\bar{Q}_v^{+}}
    \newcommand{\QVP}{Q^{\prime +}_{v^{\prime}} }
    \newcommand{\QVBP}{\bar{Q}^{\prime +}_{v^{\prime}} }
    \newcommand{\QVHZ}{\hat{Q}^{+}_v}
    \newcommand{\QVHZB}{\bar{\hat{Q}}_v{\vspace{-0.3cm}\hspace{-0.2cm}{^{+}} } }    
    \newcommand{\QVPHZB}{\bar{\hat{Q}}_{v^{\prime}}{\vspace{-0.3cm}\hspace{-0.2cm}{^{\prime +}}} }    
    \newcommand{\QVPHFB}{\bar{\hat{Q}}_{v^{\prime}}{\vspace{-0.3cm}\hspace{-0.2cm}{^{\prime -}} } }    
    \newcommand{\QVPHB}{\bar{\hat{Q}}_{v^{\prime}}{\vspace{-0.3cm}\hspace{-0.2cm}{^{\prime}} }   }
    \newcommand{\QVHF}{\hat{Q}^{-}_v}
    \newcommand{\QVHFB}{\bar{\hat{Q}}_v{\vspace{-0.3cm}\hspace{-0.2cm}{^{-}} }}    
    \newcommand{\QVH}{\hat{Q}_v}
    \newcommand{\QVHB}{\bar{\hat{Q}}_v}
    \newcommand{\VS}{v\hspace{-0.2cm}\slash} 
    \newcommand{\MQ}{m_{Q}}
    \newcommand{\MQP}{m_{Q^{\prime}}}
    \newcommand{\QVHPMB}{\bar{\hat{Q}}_v{\vspace{-0.3cm}\hspace{-0.2cm}{^{\pm}} }}    
    \newcommand{\QVHMPB}{\bar{\hat{Q}}_v{\vspace{-0.3cm}\hspace{-0.2cm}{^{\mp}} }  }  
    \newcommand{\QVHPM}{\hat{Q}^{\pm}_v}
    \newcommand{\QVHMP}{\hat{Q}^{\mp}_v}

\draft
\title{ A Consistent Calculation of Heavy Meson Decay Constants and \\
Transition Wave Functions in the Complete HQEFT}
\author{ W.Y. Wang and Y.L. Wu\footnote{ylwu@itp.ac.cn \\ Supported by Outstanding Young Scientist 
Research Found.}  }
\address{Institute of Theoretical Physics, Chinese Academy of Sciences, Beijing 100080, China } 
\maketitle

\begin{abstract} 
 Within the complete heavy quark effective field theory (HQEFT), the QCD sum rule 
approach is used to evaluate the decay constants including $1/m_Q$ corrections   
and the Isgur-Wise function and other additional important wave functions concerned at 
$1/m_Q$ for the heavy-light mesons. The number of unknown wave functions or form factors 
in HQEFT is shown to be much less than the one in the usual heavy quark effective theory (HQET). 
The values of wave functions at zero recoil are found to be consistent with the ones extracted 
from the interesting relations (which are resulted from the HQEFT) between the hadron masses and 
wave functions at zero recoil. The results for the decay constants are consistent
with the ones from full QCD sum rule and Lattice calculations. The $1/m_Q$ corrections to the scaling law 
$f_M \sim F/\sqrt{m_M}$ are found to be small in HQEFT, which demonstrates again the 
validity of $1/m_Q$ expansion in HQEFT. It is also shown that the residual momentum 
$v\cdot k$ of heavy quark within heavy-light hadrons does be around 
the binding energy $\bar{\Lambda}$ of the heavy hadrons, which turns out to be in agreement with 
the expected one in the HQEFT. Therefore such a calculation provides a consistent check on 
the HQEFT and shows that the HQEFT is more reliable than the usual HQET for describing a slightly 
off-mass shell heavy quark within hadron as the usual HQET seems to lead to 
the breakdown of $1/m_Q$ expansion in evaluating the meson decay constants. 
It is emphasized that the introduction of the `dressed heavy quark' mass is useful 
for the heavy-light mesons (Qq) with $m_Q >> \bar{\Lambda} >> m_q$, while for heavy-heavy bound states
($\psi_1\psi_2$) with masses $m_1,\ m_2 >> \bar{\Lambda}$, like bottom-charm hadrons or 
similarly for muonium in QED, one needs to treat both particles as heavy effective particles via $1/m_1$ 
and $1/m_2$ expansions and redefine the effective bound states and modified `dressed heavy quark' masses
within the HQEFT.  
\end{abstract}


\newpage

\section{Introduction}\label{int}

 It has been seen that the heavy quark effective field theory (HQEFT) with keeping both 
 quark and antiquark fields\cite{ylw} can provide a consistent description on 
 both exclusive\cite{wwy1} and inclusive\cite{wwy2,wy} decays of heavy hadrons. The extracted 
 values of $|V_{cb}|$ from both exclusive and inclusive decays have shown a good agreement. 
 The lifetime differences among bottom mesons and hadrons can also be well understood within 
 the new framework of HQEFT. Especially, it has been noticed that at zero recoil, 
 $1/m_Q$ corrections in both exclusive and inclusive decays are automatically absent without 
 imposing the equation of motion $iv\cdot D Q_v = 0$ when the physical observables are 
 presented in terms of heavy hadron masses, this is the main point that differs from 
 the usual heavy quark effective theory (HQET) or the usual heavy quark expansion. In the 
 usual framework, $1/m_Q$ corrections in the inclusive decays are absent only when the 
 inclusive decay rate is presented in terms of heavy quark mass ($m_Q$)\cite{chay} rather than the heavy 
 hadron mass ($m_H$), the situation seems to be conflict with the case in the exclusive 
 decays where the normalization is given in term of heavy hadron mass\cite{luke}. Such an inconsistency 
 in the usual HQET may be the main reason that leads to the difficulty for understanding the 
 lifetime differences among the bottom hadrons. These observations indicate 
 that the contributions and effects from antiquark fields should play a significant role for 
 understanding hadronic structures and can become important for certain physical observables.
 
 Our basic point of considerations is based on the physical picture that a heavy hadron (Qq) 
containing a single heavy quark (Q) and satisfying the condition among the heavy quark mass 
$m_Q$, the light quark mass $m_q$ and the binding energy  $\bar{\Lambda}$
\[ m_Q >> \bar{\Lambda} >> m_q  \]
may be regarded as a `dressed heavy quark' with an off-mass 
shell by an amount of binding energy $\bar{\Lambda}$. In this case, the usual quark-hadron 
duality should be extended to a `dressed-heavy quark' - `heavy-light hadron' duality. 
 Thus a more reliable heavy quark expansion for heavy-light hadron systems
 should be carried out in terms of the ``dressed heavy quark" mass defined as 
\[ \hat{m}_Q \equiv m_Q + \bar{\Lambda} = m_H - O(1/m_Q) = \lim_{m_Q \rightarrow \infty} m_H \]
 with $m_H$ the heavy hadron mass. Before proceeding, 
we would like to point out that this picture cannot naively be applied to the hadrons with 
\[ m_Q , m_q >> \bar{\Lambda} \]
This is because such hadrons must be treated as heavy-heavy hadrons, like $B_c$ meson system. 
Similarly, the muonium $(\mu e)$ must also be treated as heavy-heavy bound state system in QED. 
In general for heavy-heavy system $(\psi_1\psi_2)$ with masses $m_1, m_2 >> \bar{\Lambda}$, 
one should make expansion for both heavy particles in 
the bound state and redefine the effective fields and effective bound states. 

   It has been shown\cite{ylw,wwy1,wwy2,wy} that the new 
framework of HQEFT with keeping the antiquark fields enables us to describe such an 
off-mass shell heavy quark within hadrons as the $1/m_Q$ corrections to the transition matrix elements 
automatically vanish at zero recoil. Of particular, the number of transition form factors 
invloved in the HQEFT is much less than the one in the usual HQET. In addition, 
the new framework of HQEFT also results in some interesting relations among meson masses 
and transition formfactors at zero recoil, which enables us to extract
the important transition form factors from the known heavy meson masses.  
It is those special features that allow us to check self-consistently the validity of the 
new framework of HQEFT when applying it to the heavy-light hadron systems. This can simply be carried 
out by comparing the values of transition formfactors and 
residual momentum of heavy quark in hadron, which are extracted from the relations between the
meson masses and formfactors,  with those obtained from directly evaluating the corresponding hadronic matrix 
elements by using some reliable approaches. 
In this paper, we are going to adopt QCD sum rule approach for a practical calculation. Our paper is 
organized as follows. In Sec.\ref{hqe}, we briefly review the 
 heavy quark expansion and present a general description on form factors concerned 
 up to order $1/m_Q$, their corresponding hadronic matrix elements in heavy meson weak 
 decays and weak transitions between two heavy mesons are expilcitly defined. 
In Sec.\ref{decay}, we apply the QCD sum rules 
to two point Green functions of heavy quark currents within the framework of HQEFT, and 
calculate three form factors $F$, $G_1$ and $G_2$ as well as two composite factors $g_1$ 
and $g_2$ concerned in the decay and coupling constants 
of heavy pseudoscalar and vector mesons. It also enables us to extract a reasonable value of 
the binding energy $\bar{\Lambda}$.  In Sec.\ref{tran}, we investigate three point Green 
functions of heavy quark operators and apply the QCD sum rule to evaluate
the Isgur-Wise function and some additional wave functions appearing at the order of 
$1/m_Q$. It is also shown that the residual momentum of the heavy quark within the hadron is 
truely around the binding energy, i.e., $iv\cdot D \sim v\cdot k \approx \bar{\Lambda}$ is 
found to be a good approximation in simplifying the evaluation of the hadronic matrix elements. 
In Sec.\ref{radiative}, we add the two-loop perturbative contrinutions to the two-point 
collerator for the purpose of seeing the importance of QCD radiative corrections. 
A brief summary with some remarks is presented in Sec.\ref{sum}. 

  \section{Hadronic Matrix Elements in HQEFT}\label{hqe}

It has been shown that by decomposing the original heavy quark field in QCD 
into effective quark field and antiquark field \cite{ylw} and integrating out both the antiquark field 
and the small components of quark field, we arrive at an effective theory 
for only the large components of quark field $Q_v^+$, its effective Lagrangian has 
the following form \cite{wwy1,wwy2}
 \begin{eqnarray}
 \label{eq:Lagrangian}
   L^{(++)}_{eff} &=& L^{(0)}_{eff}+L^{(1/m_Q)}_{eff} \nonumber\\
   &=& \QVB i\DSP \QV + \frac{1}{\MQ}\QVB(i\DSC)^2\QV+O(\frac{1}{\MQ^2}), 
 \end{eqnarray} 
 where $\QV$ is the heavy quark effective field in this new framework of HQEFT, its 
momentum is $k= P_Q - m_Q v$ with $P_Q$ the momentum of the original heavy quark field 
in QCD. The operators $\DSP$ and $\DSC$ are defined as
      \begin{equation}
      \label{eq:12}
       \left\{`
        \begin{array}{l}
         \int g(x) \stackrel{\hspace{-0.1cm}\leftarrow}{D^\mu}\varphi(x) 
    \equiv -\int g(x) D^{\mu}\varphi(x), \\
           \DSP \equiv v\hspace{-0.2cm}\slash(v\cdot D), \\
           \DSC \equiv \DS-\VS (v\cdot D)
       \end{array}
         \right.
      \end{equation}
 with $v^{\mu}$ an arbitrary four-vector satisfying $v^2=1$. In general, a mass dimension 
 parameter $\Lambda$ may appear in the Lagrangian depending on the redefinition of 
heavy quark effective field\cite{wwy1}, here we take $\Lambda=0$ for the convenience
in comparison with the convention in the usual heavy quark effective theory(HQET).

 The decay constants of a heavy pseudoscalar meson $P$ and a heavy vector meson $V$ are 
defined by
 \begin{eqnarray}
 \label{eq:defdecay}
 && <0|\bar{q} \gamma_{\mu} \gamma_{5} Q|P(v)>=i f_P m_P v_{\mu} ,\nonumber\\
 && <0|\bar{q} \gamma_{\mu} Q|V(\epsilon,v)>= f_V m_V \epsilon_{\mu}
 \end{eqnarray}
 with $Q$ the original quark field in QCD. Here $\epsilon_{\mu}$ is the polarization 
vector of a vector meson. $|P(v)>$ and $|V(\epsilon,v)>$ are the pseudoscalar and vector 
meson states in QCD. 
 
 In the new framework of HQEFT, current operators composed by one 
heavy quark and one light (or heavy) quark can be expanded in terms of operators in the 
effective theory
 \begin{eqnarray}
 \label{eq:currentexp}
 \bar{q}\Gamma Q &\to& \bar{q} \Gamma \QV+\frac{1}{2m_Q} \bar{q}\Gamma\frac{1}{i \DSP}(i\DSC)^2\QV 
   +O(1/m^2_Q), \nonumber \\
 \bar{Q} \Gamma Q &\to& \QVBP \Gamma \QV+\frac{1}{2m_Q} \QVBP\Gamma\frac{1}{i \DSP}(i\DSC)^2\QV 
\nonumber\\
&&   +\frac{1}{2m_{Q'}} \QVBP (-i\stackrel{\hspace{-0.1cm}\leftarrow}{\DSC})^2
   \frac{1}{-i\stackrel{\hspace{-0.1cm}\leftarrow}{\DSP}}
   \Gamma \QV +O(1/m^2_Q). 
 \end{eqnarray}
  
 Correspondingly, it is useful to introduce an effective heavy hadron state $|H_v>$ 
for exhibiting a manifest heavy quark spin-flavor symmetry\cite{wwy1}. 
Such an effective hadron state in HQEFT is related to the hadron state $|H>$ in QCD 
via\cite{wwy1,wwy2}
   \begin{equation}
    \label{eq:statedef}    
     \frac{1}{\sqrt{m_{H^{\prime}}m_{H}}} <H^{\prime}\vert \bar{Q}^{\prime} \Gamma Q \vert H>= 
 \frac{1}{\sqrt{\bar{\Lambda}_{H^{\prime}} \bar{\Lambda}_H}} <H^{\prime}_{v^{\prime}}\vert
          J^{(++)}_{eff} e^{i\int d^4x L^{(1/{\MQ})}_{eff}}\vert H_v>.
    \end{equation}  
  with 
    \begin{equation}
    \label{eq:binding}
       \bar{\Lambda}_{H^{(')}} \equiv m_{H^{(')}}-m_{Q^{(')}}.
    \end{equation}
  The renormalization condition for $|H_v>$ is given by
   \begin{equation}
     \label{eq:stateren}
         <H_{v} \vert \QVB \gamma^{\mu} \QV \vert H_v> = 2\bar{\Lambda} v^{\mu} 
     \end{equation}
with 
\[ \bar{\Lambda} = \bar{\Lambda}_{H} - O(1/m_Q) = \lim_{m_{Q}\to \infty} \bar{\Lambda}_H \]
being a heavy flavor independent binding energy that reflects the 
effects of the light degrees of freedom in the heavy hadron. Obviously, the above 
normalization condition preserves spin-flavor symmetry. We would like to address again that 
above renormalization is only applicable for the heavy-light hadrons with $m_Q >> \bar{\Lambda} >> m_q$. 
For heavy-heavy bound state system, i.e. $m_1, m_2 >> \bar{\Lambda}$,  such as bottom-charm meson
$B_c$ and muonium $(\mu e)$, one needs to redefine the effective bound states by considering $1/m_Q$
expansion for both heavy particles.
 
 In the heavy quark expansion, $1/m_Q$ order corrections to the hadronic matrix elements 
 arise from not only the current expansion (\ref{eq:currentexp}) but also the effective 
 Lagrangian, i.e., insertion of $L^{(1/m_Q)}_{eff}$ into the matrix elements. By including 
 all these corrections, we arrive at the following results for the hadronic matrix elements 
of the currents in (\ref{eq:currentexp}) 
 \begin{eqnarray}
 \label{eq:matrixexp}
 &&\sqrt{\frac{\bar{\Lambda}_M}{m_M}} <0|\bar{q}\Gamma Q|M> \to 
  <0|\bar{q} \Gamma \QV|M_v>-\frac{1}{2m_Q} <0|\bar{q}\Gamma\frac{1}{i \DSP}(i\DSC)^2\QV |M_v>
   +O(1/m^2_Q), \nonumber \\
&& \sqrt{\frac{\bar{\Lambda}_{M'} \bar{\Lambda}_{M} }{m_{M'} m_M}} <M'|\bar{Q}' \Gamma Q|M> \to 
 <M'_{v'}|\QVBP \Gamma \QV|M_v>-\frac{1}{2m_Q} <M'_{v'}|\QVBP\Gamma\frac{1}{i \DSP}(i\DSC)^2\QV |M_v>
\nonumber\\
&&\hspace{2cm}  -\frac{1}{2m_{Q'}} <M'_{v'}|\QVBP (-i\stackrel{\hspace{-0.1cm}\leftarrow}{\DSC})^2
   \frac{1}{-i\stackrel{\hspace{-0.1cm}\leftarrow}{\DSP}}
   \Gamma \QV|M_v> +O(1/m^2_Q). 
 \end{eqnarray}

The relevant hadronic matrix elements may be parameterized as follows 
 \begin{eqnarray}
 \label{eq:parametrization}
&& <0|\bar{q}\Gamma\QV |M_v>=\frac{F}{2} Tr[\Gamma {\cal M}],\nonumber \\
&& <0|\bar{q}\Gamma \frac{-1}{i v\cdot D}P_{+} (\DC)^2 \QV |M_v>=-F G_1 Tr[\Gamma {\cal M}],\nonumber \\
&& <0|\bar{q}\Gamma \frac{-1}{i v\cdot D}P_{+} \frac{i}{2} \sigma_{\alpha\beta} F^{\alpha\beta}
     \QV |M_v>=
     2F G_2 Tr[i \sigma_{\alpha \beta} \Gamma P_{+} \frac{i}{2} \sigma^{\alpha \beta} {\cal M} ]
     =-2 F G_2 d_M Tr[\Gamma {\cal M} ],   \nonumber \\
&& <M^{\prime}_{v^{\prime}}\vert \QVBP\Gamma \QV \vert M_v> =-\xi(y)
     Tr[\bar{\cal M}^{\prime}\Gamma {\cal M}], \nonumber   \\ 
&& <M^{\prime}_{v^{\prime}}\vert \QVBP\Gamma \frac{-1}{iv\cdot D} P_{+} \DC^2 
      \QV \vert M_v> =-\kappa_1(y) \frac{1}{\bar{\Lambda}} Tr[\bar{\cal M}^{\prime}\Gamma {\cal M}],
   \nonumber      \\
&& <M^{\prime}_{v^{\prime}}\vert \QVBP\Gamma \frac{-1}{iv\cdot D} P_{+} \frac{i}{2}
      \sigma_{\alpha\beta}F^{\alpha\beta}\QV \vert M_v> =\frac{1}{\bar{\Lambda}}
Tr[\kappa_{\alpha\beta}(v,v^{\prime})
      \bar{\cal M}^{\prime}\Gamma P_{+}\frac{i}{2}\sigma^{\alpha\beta}{\cal M}],
\end{eqnarray}
 where $y=v\cdot v'$. ${\cal M}(v)$ is the spin wave function
   \begin{eqnarray}
     \label{eq:spinwave}
       {\cal M}(v)=\sqrt{\bar{\Lambda}}P_{+}
         \left\{
           \begin{array}{cl}
              -\gamma^{5}, & \mbox{for pseudoscalar meson} \; \; P\\
              \epsilon\hspace{-0.15cm}\slash, & \mbox{for vector meson} \; \; V
           \end{array}
         \right.
     \end{eqnarray}
 and 
   \begin{eqnarray}
     \label{eq:dm}
       d_M=
         \left\{
           \begin{array}{cl}
              3, & \mbox{for pseudoscalar meson} \; \; P\\
              -1, & \mbox{for vector meson} \; \; V ,
           \end{array}
         \right.
     \end{eqnarray}
 where $F$, $G_1$ and $G_2$ are constants, and $\xi(y)$ and $\kappa_1(y)$ the
 Lorentz scalar functions. Actually, $\xi(y)$ is the well-known Isgur-Wise functions.
 The Lorentz tensor $\kappa_{\alpha \beta}(v,v')$ can be decomposed into 
 \begin{eqnarray}
 \label{eq:kappadecomp}
   \kappa_{\alpha \beta}(v,v')=i \kappa_2(y) \sigma_{\alpha \beta} +\kappa_3(y)
   (v'_{\alpha} \gamma_{\beta}-v'_{\beta} \gamma_{\alpha})
 \end{eqnarray}
 with $\kappa_2(y)$ and $\kappa_3(y)$ being the Lorentz scalar functions. 
 
 Combining (\ref{eq:defdecay}), (\ref{eq:stateren}) and (\ref{eq:parametrization}) we have
 \begin{eqnarray}
 \label{decayrelation}
 f_M=\sqrt{\frac{\bar{\Lambda}}{\bar{\Lambda}_M m_M} } F \{ 1+\frac{1}{m_Q}(G_1+2d_M G_2)\}\, .
 \end{eqnarray}
 Thus the ratio between the vector and pseudoscalar meson constants is given by
 \begin{eqnarray}
 \label{eq:ratio}
 \frac{f_V m^{1/2}_V}{f_P m^{1/2}_P}=(\frac{\bar{\Lambda}_P}{\bar{\Lambda}_V})^{1/2} (1-\frac{8}{m_Q} G_2).
 \end{eqnarray}

 As is known, the normalization of the Isgur-Wise function at zero recoil point is given
by $\xi(1)=1$. The additional wave functions $\kappa_1(y)$, $\kappa_2(y)$ and $\kappa_3(y)$ 
characterize the next-to-leading order symmetry-breaking corrections to $\xi$. 
From (\ref{eq:parametrization}) and (\ref{eq:kappadecomp}), it is easily seen that 
the hadronic matrix element at zero recoil is irrelevant to $\kappa_3(1)$. While 
$\kappa_1(1)$ and $\kappa_2(1)$ are found to be related to the meson masses\cite{wwy1,wwy2} 
 \begin{eqnarray}
   \label{massrelation}
       \bar{\Lambda}_{M}=m_M-m_Q=
          \bar{\Lambda}-(\frac{1}{m_Q}-\frac{\bar{\Lambda}}{2m^2_Q})
          (\kappa_1(1)+d_M \kappa_2(1)).
  \end{eqnarray}
  
 It is easy to check that the leading order contributions to the hadronic matrix elements 
in the HQEFT, which are characterized by the decay constant $F$ and the Isgur-Wise 
function $\xi$ defined in (\ref{eq:parametrization}),  are the same as the ones 
in the usual HQET. 
Nevertheless, to the next-to-leading order, differences occur between the two frameworks. In 
 the usual HQET, the transition matrix elements between two heavy mesons are parameterized 
 by six functions denoted as $\xi_i$ and $\chi_i$ ($i=1,2,3$). Here $\xi_1$, $\xi_2$ and 
 $\xi_3$ arise from the current expansion, and $\chi_1$, $\chi_2$ and $\chi_3$ from the 
 insertion of the effective Lagrangian into the hadronic matrix elements. All these six 
quantities are functions of the recoil parameter $y=v \cdot v'$. In addition, there are 
two more parameters $\lambda_1$ and $\lambda_2$ which appear in the $1/m_Q$ order corrections 
to heavy meson masses. Unlikely, in the new framework of HQEFT, it has been noticed that
 the evaluation of the hadronic matrix elements is greatly simplified when the antiquark 
 contributions are included\cite{ylw,wwy1,wwy2}. As a consequence, all
 transition matrix elements concerned at $1/m_Q$ order can be characterized by only three 
wave functions $\kappa_1(y)$, $\kappa_2(y)$ and $\kappa_3(y)$. Furthermore, the $1/m_Q$ order 
corrections to meson masses were found to be naturally related to their values at zero recoil, 
i.e., $\kappa_1(1)$ and $\kappa_2(1)$. In other words, instead of the six wave functions 
$\xi_i(y)$ and $\chi_i(y)$ (i=1,2,3) and two parameters $\lambda_1$ and $\lambda_2$ in the 
usual HQET,  we only need to evaluate three wave functions $\kappa_i$ ($i=1,2,3$) in the new 
framework of HQEFT up to the order $1/m_Q$. At zero recoil, only two parameters $\kappa_1(1)$ 
and $\kappa_2(1)$ are relevant. Similar comments hold for the corrections to meson decay 
constants. In the HQEFT one encounters only two constants $G_1$ and $G_2$ at $1/m_Q$ order. 
 
Within the new framework of HQEFT, two important parameters 
$\kappa_1(1)$ and $\kappa_2(1)$ have been extracted from meson mass spectrum in 
refs.\cite{wwy1}. To have an independent check for the HQEFT, it would be very useful 
to evaluate these two wave functions directly by a field-theoretical method within the 
framework of HQEFT. In the following sections, we shall present an QCD sum rule study 
for these form factors. 
 
 \section{QCD sum rule Calculation of $F$, $G_1$ and $G_2$}\label{decay}

 QCD sum rule approach has widely been used to calculate hadronic matrix elements in QCD and 
 has also been applied to effective theories of QCD. It turns out to be a powerful analytic 
 approach to estimate non-perturbative effects. The basic idea 
 of QCD sum rule formalism is to study the analytic properties of correlation functions, 
 and to treat the bound state problems in QCD from quark-hadron duality considerations. 
 So that one could start at short distance physics and moves to large distance physics
 where confinement effects become important and resonances emerge as a reflection of confinement. 
 Here we shall briefly outline the main steps of sum rule treatments. In general, 
 the fixed point gauge for the background fields is used in calculating Feynman diagrams.
  
 In order to evaluate the parameters $F$, $G_1$ and $G_2$ involved in the meson decay 
constants, we consider the following two-point correlator 
 \begin{eqnarray}
\label{eq:2point}
 \Pi (\omega)= i \int d^4x e^{i P\cdot x} <0|(\bar{q} \Gamma_{M'} Q)_{(x)}, (\bar{Q} \Gamma_{M} q)_{(0)}|0>,
\end{eqnarray}
 where $\Gamma_M$ has appropriate Lorentz structure so that the two currents in (\ref{eq:2point}) 
 interpolate the heavy meson of interest. It is convenient to choose \cite{neu1076}
   \begin{eqnarray}
     \label{eq:GammaM}
       \Gamma_M=
         \left\{
           \begin{array}{cl}
              -i \gamma^{5}, & \mbox{pseudoscalar meson} \; \; P\\
              \gamma_{\mu}-v_{\mu}, & \mbox{vector meson} \; \; V
           \end{array}
         \right.
  \end{eqnarray} 
The total external momentum in (\ref{eq:2point}) is $P=m_Q v+k$ with 
the momentum $k$ in (\ref{eq:2point}) being the residual momentum of the heavy quark.
 The correlator $\Pi$ is an analytic function of $2v\cdot k+k^2/m_Q$ with discontinuities
for its positive values. Here $k^\mu=k^{\mu}_T+k^{\mu}_L$ with $k^\mu_T$ and 
$k^\mu_L=(v\cdot k) v^\mu$ being the transverse 
and longitudinal part of $k$ separately. Particularly, under the definition 
$\omega \equiv 2v\cdot k+\frac{k^2_T}{m_Q}$, one has 
 \begin{eqnarray}
  \label{eq:defomeQ}
  2v\cdot k+\frac{k^2}{m_Q} = \omega+\frac{\omega^2}{4 m_Q} + O(1/m^2_Q).
 \end{eqnarray}
In eq.(\ref{eq:2point}) we have represented $\Pi$ as an analytic function of the 
variable $\omega$.

 Phenomenologically, the two-point function $\Pi(\omega)$ can be written as the sum of three 
parts: a pole contribution from the ground state mesons associated with the heavy-light 
currents; a dispersion integral over a physical spectral function; and subtraction terms, 
namely,
 \begin{eqnarray}
  \label{eq:2pointphen}
  \Pi_{phen}(\omega)&=&
 -(\sum_{pole}) \frac{<0|\bar{q} \Gamma_M Q|M><M|\bar{Q} \Gamma_M q|0>}{P^2-m^2_M+i\epsilon} \nonumber\\
 && +\int^{\infty}_{\omega_c} d\nu \frac{\rho_{phys}(\nu)}{\nu-\omega-i\epsilon }
 +subtractions,
 \end{eqnarray}
 where the two matrix elements should be expanded by using eq.(\ref{eq:matrixexp}). 
$(\sum_{pole})$ means summation over polarization for vector mesons. 
 
 On the other hand, using the Feynman rules of the HQEFT, the correlation function in HQEFT 
 is evaluated perturbatively in the deep Euclidean region 
 ($\omega \ll \bar{\Lambda}$) via
 \begin{eqnarray}
\label{eq:2pointinhqeft}
 \Pi (\omega)&=& i \int d^4x e^{i k\cdot x}  \{ <0|(\bar{q} \Gamma_{M'} \QV)_{(x)}, 
      (\QVB \Gamma_{M} q)_{(0)} |0> \nonumber \\
   &+&\frac{1}{2 m_Q} <0|(\bar{q} \Gamma_{M'} \frac{1}{i \DSP} (i\DSC)^2 \QV)_{(x)},
      (\QVB \Gamma_{M} q)_{(0)} |0> \nonumber \\
   &+&\frac{1}{2 m_Q} <0|(\bar{q} \Gamma_{M'} \QV)_{(x)}, 
      (\QVB (-i\stackrel{\hspace{-0.1cm}\leftarrow}{\DSC})^2
      \frac{1}{-i\stackrel{\hspace{-0.1cm}\leftarrow}{\DSP}}
\Gamma_{M} q)_{(0)} |0> \} \nonumber \\
   &+&O(1/m^2_Q)
\end{eqnarray}
  In lower Euclidean region, the 
 non-perturbative contributions become important. In QCD sum rule analysis, the two-point 
 correlator receives contributions not only from the pure perturbative ones but 
 also the ones from condensates, which characterizes non-vanishing vacuum expectation 
 values of local operators in the operator product expansion(OPE). It is thought that 
 by including these condensates, the QCD confinement effects may be accounted for at the 
 transition from perturbative region to non-perturbative region. 
 Writing the perturbative contributions as the form of an integral over a theoretic spectral function, 
 the theoretical result for the two-point correlator (\ref{eq:2point}) becomes
 \begin{eqnarray}
 \label{2pointtheo}
  \Pi_{theor}(\omega)=\int d\nu \frac{\rho_{pert}(\nu)}{(\nu-\omega-i\epsilon)}
  + \Pi_{NP}+subtractions.
 \end{eqnarray}

 A basic assumption in QCD sum rule is the quark-hadron duality. Due to this duality one
 can model the contributions of higher resonance states by the perturbative continuum 
 starting at a threshold energy $\omega_c$. In other words, we assume 
 $\rho_{phys}=\rho_{pert}$.  Equating the phenomenological side and the theoretical side, 
 up to the order of $1/m_Q$ one arrives at 
 \begin{eqnarray}
 \label{eq:phethe}
 & 2 Tr[\Gamma_M P_{+} \Gamma_M] 
\frac{F^2}{4} 
  [1+\frac{2}{m_Q} (G_1+2 d_m G_2) ] \frac{\bar{\Lambda}}{\bar{\Lambda}_M} 
  \frac{m_M}{m_Q} \frac{1}{\omega-\omega_M+i\epsilon} (1-\frac{\omega+\omega_M}{4 m_Q}) 
     \nonumber\\
 &= \int^{\omega_c}_{0} d\nu \frac{\rho_{pert}(\nu)}{(\nu-\omega-i\epsilon)}
     + \Pi_{NP}+subtractions
 \end{eqnarray}
 with $\omega_M \equiv 2\bar{\Lambda}_M$. 
 In deriving (\ref{eq:phethe}) we have used (\ref{eq:parametrization}) as well as the relations
 \begin{eqnarray}
 \label{eq:diracrel}
   Tr[\bar{\Gamma}_M {\cal M}(v)] Tr[\bar{{\cal M}}(v) \Gamma_M]=-2\bar{\Lambda} 
  Tr[\bar{\Gamma}_M P_{+} \Gamma_M]
 \end{eqnarray}
 and
 \begin{eqnarray}
 \label{eq:denuome}
   \frac{P^2-m^2_M}{m_Q}=(\omega-\omega_M)[1+\frac{\omega+\omega_M}{4 m_Q}+O(1/m^2_Q)].
 \end{eqnarray}
 
 The relevant Feynman diagrams are plotted in Fig.1. Fig.1(a) is the lowest order 
 perturbative diagram. For the non-perturbative effects, it is sufficient in the present case 
 to consider only the contributions of the quark condensate, the gluon condensate and the 
mixed quark-gluon condensate, which have values ($\alpha_s \equiv g^2_s/4\pi$)
 \begin{eqnarray}
   &&<\bar{q} q>\approx -(230) \; \mbox{MeV}^3; \nonumber\\
   &&i<\bar{q} \sigma_{\alpha \beta} F^{\alpha \beta} q>\approx -0.8 <\bar{q} q>; \nonumber\\
   &&\alpha_s <FF>\equiv \alpha_s <F^a_{\alpha\beta} F^{\alpha \beta}_a>\approx 0.04 \;\mbox{GeV}^4 .
 \end{eqnarray}
 We only keep terms up to order $\alpha_s$ for the condensates. Neglecting the light quark 
mass, we obtain the following results at the renormalization scale $\mu \sim 2 \bar{\Lambda}$
 \begin{eqnarray}
 \label{eq:beforeBT}
   &&F^2 [1+\frac{2}{m_Q}(G_1+2 d_M G_2)-\frac{1}{m_Q }
   \frac{(\omega^{(1)}_M+d_M \omega^{(2)}_M )}{2 \bar{\Lambda}}+\frac{\bar{\Lambda}}{m_Q}  ]
   [ \frac{1}{4 m_Q}+\frac{1}{\omega_M-\omega-i \epsilon} (1-\frac{\omega_M}{2 m_Q}) ]
   = \nonumber \\
  && \frac{3}{8\pi^2} \int^{\omega_c}_{0} d \nu 
   [\frac{1}{4 m_Q}+\frac{1}{\nu-\omega-i\epsilon} ]
   ({\nu}^2-\frac{3 {\nu}^3 }{4 m_Q})
    +<\bar{q}q> [\frac{1}{\omega}-\frac{1}{4m_Q} 
   +\frac{\alpha}{\pi} (\frac{4}{3\omega}-\frac{9-4 d_M}{9m_Q} )  ] \nonumber\\
   && +i<\bar{q}\sigma_{\alpha\beta} F^{\alpha\beta} q> [ \frac{1}{2\omega^3}
   -\frac{1}{m_Q \omega^2} (\frac{3}{8}-\frac{d_M}{12})+\frac{\alpha}{\pi} 
   (\frac{2}{\omega^3}+\frac{163+147 d_M}{576m_Q \omega^2} ) ] \nonumber\\
   &&-\frac{\alpha}{\pi} <FF> [\frac{1}{24 \omega^2}-\frac{15-4 d_M}{96 m_Q \omega}] +subtractions.
 \end{eqnarray}

 In order to improve the convergence and suppress the importances of higher-resonance states, 
 we apply the Borel operator
 \begin{eqnarray}
 \label{eq:defBT}
   \hat{B}^{(\omega)}_{T}\equiv T \; {\mbox{lim}_{n->\infty}}_{-\omega->\infty} 
   \frac{ \omega^n}{\Gamma(n)} (-\frac{d}{d\omega})^n 
 \end{eqnarray}
 to both sides of (\ref{eq:beforeBT}) with $ T=\frac{-\omega}{n}$ being held fixed. 
 In the dispersion integral, this Borel transformation yields an 
 exponential damping factor which effectively suppresses high-resonance contributions. On the 
 non-perturbative terms, Borel transformation enhances the importance of low dimension 
 condensates. Furthermore, the subtraction terms in (\ref{eq:beforeBT}) may also be eliminated by this 
 Borel transformation. 
 
 The resulting Borel transformed sum rule reads 
 \begin{eqnarray}
 \label{eq:decaySR}
   &&F^2 [1+\frac{2}{m_Q}(G_1+2 d_M G_2)-\frac{1}{2 m_Q \bar{\Lambda}}
   (\omega^{(1)}_M+d_M \omega^{(2)}_M )+\frac{\bar{\Lambda}}{m_Q}-\frac{\omega_M}{2 m_Q} ]
   e^{-\omega_M/T}      =\nonumber\\
  && \frac{3}{8\pi^2} \int^{\omega_c}_{0} d \nu e^{-\nu/T}
    (\nu^2-\frac{3 {\nu}^3 }{4 m_Q}) \nonumber \\
&& -<\bar{q}q>(1+\frac{4\alpha_s}{3\pi} )-i<\bar{q}\sigma_{\alpha\beta} F^{\alpha\beta} q> 
 [\frac{1}{4 T^2} +\frac{1}{m_Q T}(\frac{3}{8}-\frac{d_M}{12}) \nonumber\\
&& +\frac{\alpha_s}{\pi} (\frac{1}{T^2}-\frac{163+147 d_M}{576 m_Q T})] 
-\frac{\alpha_s}{\pi} <FF>(\frac{1}{24 T}+\frac{15-4d_M}{96m_Q}).
 \end{eqnarray}

 Besides the $1/m_Q$ corrections shown explicitly in (\ref{eq:decaySR}), the pole energy 
$\omega_M$ and the threshold energy $\omega_c$ also receive $1/m_Q$ corrections. 
We may write them as
 \begin{eqnarray}
  \label{eq:defomegacor}
  \omega_M&\equiv&2\bar{\Lambda}_M=\omega^{(0)}_M+\frac{1}{m_Q}(\omega^{(1)}_M+d_M \omega^{(2)}_M),
  \nonumber\\
  \omega_c&=&\omega_0+\frac{1}{m_Q}(\omega_1+d_M \omega_2).
 \end{eqnarray}
With these formulae, eqs.(\ref{decayrelation}) and (\ref{eq:ratio}) can be rewritten as 
 \begin{eqnarray}
 \label{decayrelationwithg}
 f_M=\frac{F}{\sqrt{m_M}} \{ 1+\frac{1}{m_Q}(g_1+2d_M g_2)\}
 \end{eqnarray}
 and 
  \begin{eqnarray}
 \label{eq:ratiowithg}
 \frac{f_V m^{1/2}_V}{f_P m^{1/2}_P}=1-\frac{8}{m_Q} g_2 
 \end{eqnarray}
 with $g_1$ and $g_2$ being two composite factors defined as 
 \begin{eqnarray}
  \label{eq:gdef}
   g_1 &\equiv& G_1-\frac{\omega^{(1)}_M}{4\bar{\Lambda}}, \nonumber \\
   g_2 &\equiv& G_2-\frac{\omega^{(2)}_M}{8\bar{\Lambda}} .
  \end{eqnarray}

 Now we first consider the sum rule (\ref{eq:decaySR}) at leading order. For the leading 
 terms, the results are the same as the ones in HQET, because there is no difference 
 between the HQEFT and the usual HQET in the limit $m_Q\to \infty$. 
 We shall not repeat the analysis for the leading order sum rule analysis, 
 our numerical results are presented in Fig.4, where we 
 have used $\alpha_s(2\bar{\Lambda})\simeq 0.34$. From the stability of the curves, we are led
 to the following solutions for the parameters 


 \begin{eqnarray}
  \label{eq:decval1}
   \omega_0&=&1.8\pm 0.3 \; \mbox{GeV},\nonumber\\
   \frac{\omega^{(0)}_M}{2}&=&\bar{\Lambda}=0.53\pm 0.08 \; \mbox{GeV},\nonumber\\
   F&=&0.30\pm 0.06 \; \mbox{GeV}^{3/2}.
  \end{eqnarray}

 We now proceed to the next-to-leading order sum rule analysis in (\ref{eq:decaySR}). Putting 
(\ref{eq:defomegacor}) into (\ref{eq:decaySR}), and expanding (\ref{eq:decaySR}) in $1/m_Q$, 
we obtain, at $1/m_Q$ order, two sum rule formulae which are relevant to the spin-symmetry 
conserving and violating corrections, respectively. These two sorts of corrections are easy 
to be distinguished because the latter is proportional to $d_M$. To find out the solutions,
 it is useful to first evaluate the quantities $\omega_1$ and $\omega_2$ by requiring 
 optimal stability of $\omega^{(1)}_{M}$ and $\omega^{(2)}_{M}$ with respect 
 to the Borel parameter $T$ in the allowed sum rule windows. Using the central values of 
$\omega_0$, $\bar{\Lambda}$ and $F$ in (\ref{eq:decval1}), we then obtain for $\omega_i$, 
$\omega^{(i)}_{M}$, $G_i$ and $g_i$ the numerical 
results plotted in Figs.5-10. When the Borel parameter $T$ takes the reliable values 
$T= 1 \pm 0.2$ GeV, one can read off the following solutions


   \begin{eqnarray}
  \label{eq:decval2}
   \omega_1 &=& 1.5\pm 0.2 \; \mbox{GeV}^2, \nonumber\\
   \omega^{(1)}_M &=& 0.86\pm 0.10 \; \mbox{GeV}^2, \nonumber\\
   G_1&=&0.95 \pm 0.15 \; \mbox{GeV},  \nonumber \\
   g_1&=&0.54 \pm 0.12 \; \mbox{GeV},  \nonumber \\
   \omega_2 &=& -0.15\pm 0.05 \;  \mbox{GeV}^2, \nonumber\\
   \omega^{(2)}_M &=& -0.16\pm 0.03 \; \mbox{GeV}^2, \nonumber\\
   G_2 &=& -0.09\pm 0.03 \; \mbox{GeV}, \nonumber \\
   g_2 &=& -0.06\pm 0.02 \; \mbox{GeV}.
  \end{eqnarray}

 In the usual HQET\cite{neu1076} two parameters $G_1$ and $G_2$ were defined for the 
  $1/m_Q$ corrections to decay constants. 
The sum rule calculation in \cite{neu1076} yielded an unexpectedly large value for 
 $|G_1|$: $G_1\approx -4 \bar{\Lambda} \approx -2.0$GeV, which leads to the
breakdown of $1/m_Q$ expansion for decay constants. In the new framework of HQEFT, however, 
as can be seen from eqs.(\ref{decayrelationwithg}) and (\ref{eq:ratiowithg}), $1/m_Q$ 
corrections to the physical decay constants $f_M$ and the ratio are actually characterized 
by the composite factors $g_i$. Though $G_1$ in eq.(\ref{eq:decval2}) is large, $\frac{|g_1|}{m_Q}$ 
remains small enough so that the $1/m_Q$ expansion in the new framework of HQEFT 
appears to be more reliable.

 When taking the typical values for the quark masses $m_b=4.8\pm 0.10$GeV and $m_c=1.35\pm 0.10$GeV,
 we obtain from (\ref{eq:decval1}) and (\ref{eq:decval2}) the following 
values for the decay constants of bottom and charm meson without including QCD corrections
caused by the running energy scale from $\mu \simeq m_b$ to $\mu \simeq 2\bar{\Lambda}$
 \begin{eqnarray}
  \label{eq:decval3}
   &f_B(2 \bar{\Lambda})=0.135\pm 0.035  \; \mbox{GeV}, \hspace{2cm}
   &f_{B^{\ast}}(2 \bar{\Lambda})=0.147\pm 0.034 \;  \mbox{GeV},\nonumber\\
   &f_D(2 \bar{\Lambda})=0.246\pm 0.097  \; \mbox{GeV}, \hspace{2cm}
   &f_{D^{\ast}}(2 \bar{\Lambda})=0.308 \pm 0.091 \;  \mbox{GeV}.
  \end{eqnarray}
 Note that all the quantities $\omega^{(0)}_M$, $\omega_0$, $F$, 
$\omega^{(i)}_M$, $\omega_{i}$, $G_{i}$ ($i=1,2$) and $g_i$ obtained by QCD sum rules are scale 
dependent. And the results in (\ref{eq:decval1}) and (\ref{eq:decval2}) are corresponding to 
the values with the renormalization scale at $\mu \simeq 2\bar{\Lambda}\approx 1$GeV. 
So do the decay constants in (\ref{eq:decval3}). 

  From the results given in (\ref{eq:decval1}) and (\ref{eq:decval2}), we arrive at 
the following relations from eq.(\ref{eq:ratiowithg}) 
 \begin{eqnarray}
 \label{eq:rat1}
   \frac{f_{B^{\ast}} m^{1/2}_{B^{\ast}} }{f_B m^{1/2}_B } \approx 1.10 \pm 0.03, \\
 \label{eq:rat2}  
   \frac{f_{D^{\ast}} m^{1/2}_{D^{\ast}} }{f_D m^{1/2}_D } \approx 1.31 \pm 0.07.
 \end{eqnarray}
 These two ratios agrees well with the lattice calculations \cite{lattice} which lead to 
$1.12\pm 0.05$ and $1.34\pm 0.07$, respectively. 

The QCD corrections may be considered via the renormalization of the current 
$J=\bar{q}\Gamma Q$ \cite{neu1076,ylw1}. 
In renormalization-group-improved perturbation theory, up to the next-to-leading order 
the values of the decay and coupling constants at energy
scale $\mu = m_Q$ are given by  
 \begin{eqnarray}
 \label{eq:run1}
 f_M(m_Q) &\approx& \sqrt{\frac{\bar{\Lambda}}{m_M \bar{\Lambda}_M }} F(m_Q)
   [1+d_M \frac{\alpha_s(m_Q)}{6 \pi }] [1+\frac{1}{m_Q} (G_1+2d_M G_2)] \nonumber \\
  &=&  \frac{F(m_Q)}{\sqrt{m_M}} [1+d_M \frac{\alpha_s(m_Q)}{6 \pi }] 
  [1+\frac{1}{m_Q} (g_1+2d_M g_2)], \\
 \label{eq:run2}
 F(m_Q) &\approx & [\frac{\alpha_s(2\bar{\Lambda})}{\alpha_s(m_Q)}]^{6/\beta} 
  \{ 1-0.894 \frac{\alpha_s(m_Q)-\alpha_s(2\bar{\Lambda})}{\pi} 
  -\frac{ \alpha_s(m_Q) }{2\pi} \} F(2\bar{\Lambda})
 \end{eqnarray}
 with $\beta=33-2 n_F$, $M=B, D, B^{\ast}, D^{\ast}$ and $m_Q = m_b, m_c$. From the results of 
$f_M(2 \bar{\Lambda})$ given in (\ref{eq:decval3}), we then have 
\begin{eqnarray}
  \label{eq:decval3QCD}
   &f_B(m_b)=0.159\pm 0.042  \; \mbox{GeV}, \hspace{2cm}
   &f_{B^{\ast}}(m_b)=0.166\pm 0.038 \;  \mbox{GeV},\nonumber\\
   &f_D(m_c)=0.251 \pm 0.099  \; \mbox{GeV}, \hspace{2cm}
   &f_{D^{\ast}}(m_c)=0.293\pm 0.086 \;  \mbox{GeV}.
  \end{eqnarray}
which is consistent with the experimental upper limit $f_B(m_B)<200$ MeV and 
$f_D(m_D)<290 MeV$ and also some theoretical upper bounds\cite{ylw2}. The results also
agree with the lattice calculations\cite{LQCD1,LQCD2,LQCD3,LQCD4,LQCD5} 
and with full QCD calculations\cite{sch}.
 
 The ratio in eq.(\ref{eq:ratio}) is now modified to be
 \begin{eqnarray}
 \label{eq:finalratio}
 \frac{f_V m^{1/2}_V}{f_P m^{1/2}_P}&=&(\frac{\bar{\Lambda}_P}{\bar{\Lambda}_V})^{1/2} 
 (1-\frac{2\alpha_s(m_Q)}{3\pi}) (1-\frac{8}{m_Q} G_2)  \nonumber \\
 &=& (1-\frac{2\alpha_s(m_Q)}{3\pi}) (1-\frac{8}{m_Q} g_2), 
 \end{eqnarray}
 which yields
  \begin{eqnarray}
 \label{eq:finalrat1}
   \frac{f_{B^{\ast}} m^{1/2}_{B^{\ast}} }{f_B m^{1/2}_B } \approx 1.05 \pm 0.03, \\
 \label{eq:finalrat2}  
   \frac{f_{D^{\ast}} m^{1/2}_{D^{\ast}} }{f_D m^{1/2}_D } \approx 1.22 \pm 0.07.
 \end{eqnarray}

 \section{QCD sum rule evaluation on wave functions $\xi$ and $\kappa_i$}\label{tran}

  It is easily seen that to the leading order of heavy quark expansion, $m_Q \rightarrow \infty$, 
there is no difference between the HQEFT and the usual HQET, thus the procedure of calculating 
  Isgur-Wise function $\xi(y)$ is the same as in 
  the usual HQET.
The main task in this 
 section is to evaluate the two additional wave functions $\kappa_1(y)$ and $\kappa_2(y)$ 
 involved at $1/m_Q$ order. For completeness, we also briefly outline the 
  calculation of the Isgur-Wise functions $\xi(y)$ in HQEFT. 
 
  To evaluate the wave functions, we need to consider three point Green functions of the 
relevant operators. For the Isgur-Wise function $\xi(y)$, it relates the following 
three-point correlation function
at leading order of $1/m_Q$
  \begin{eqnarray}
   \label{eq:3point}
   \Xi(\omega,\omega')&=& \int d^4x d^4y e^{i(P'\cdot x-P\cdot y)} <0|(\bar{q} \bar{\Gamma}_{M'} 
   Q)_{(x)},
(\bar{Q} \Gamma Q )_{(0)}, (\bar{Q} \Gamma_{M} q)_{(y)}|0> \nonumber \\
   &=&\int d^4x d^4y e^{i(k'\cdot x-k\cdot y)} <0|(\bar{q} \bar{\Gamma}_{M'} \QVP)_{(x)},
      (\QVBP \Gamma \QV )_{(0)}, (\QVB \Gamma_{M} q)_{(y)}|0> \nonumber \\
   &+&O(1/m_Q),
  \end{eqnarray}
  where the Dirac structure $\Gamma$ of the heavy-heavy current can in principle be arbitrary. 
  For the present case, we take $\Gamma=\gamma^{\mu}$. As $\Xi(\omega,\omega')$ is an analytic 
  function in $\omega=2v \cdot k+O(1/m_Q)$ and $\omega'=2 v'\cdot k'+O(1/m_Q)$ with discontinuities on 
  the positive real axis, one can write the phenomenological representation as 
  \begin{eqnarray}
  \label{eq:IWphen}
 &&\Xi_{phen}= -(\sum_{pole}) \frac{<0|\bar{q} \bar{\Gamma}_M \QVP |M'_{v'}><M'_{v'}|\QVBP 
 \Gamma \QV|M_v> <M_v|\QVB \Gamma_M q|0>}{\bar{\Lambda}_{M} \bar{\Lambda}_{M'} 
 ( \omega_M-\omega-i\epsilon) (\omega'_M-\omega'-i\epsilon)} \frac{m_M m_{M'}}{m^2_Q} \nonumber \\
 &&+\int d\nu d\nu' \frac{\rho_{phys}(\nu,\nu')}{(\nu-\omega-i\epsilon)(\nu'-\omega'-i\epsilon)}
  +subtractions.
 \end{eqnarray}
 The first term in (\ref{eq:IWphen}) is a double-pole contribution and the second represents
  the higher resonance contributions in the form of a double dispersion integral over 
 physical intermediate states. Parameterizing the three matrix elements in (\ref{eq:IWphen}) 
by using (\ref{eq:parametrization}) and noticing the following relation
   \begin{eqnarray}
   \label{eq:diracsimp}
      Tr[\bar{\Gamma}_{M'} {\cal M}(v') ] Tr[\bar{{\cal M}}(v') \Gamma {\cal M}(v)]
     Tr[\bar{ {\cal M}}(v) \Gamma_M]=4 {\bar{\Lambda}}^2 Tr[\bar{\Gamma}_{M'} P'_{+} \Gamma P_{+} \Gamma_M],
   \end{eqnarray}
  the double-pole term becomes 
  \begin{eqnarray}
  \label{eq:IWpole}
  \Xi_{pole}=\frac{Tr[\bar{\Gamma}_M P'_{+} \Gamma P_{+} \Gamma_M]}{(\omega_M-\omega-i\epsilon)
  (\omega'_M-\omega'-i\epsilon)} \frac{\bar{\Lambda}^2}{\bar{\Lambda}_M \bar{\Lambda}_{M'} } 
  \frac{m_M m_{M'}}{m^2_Q} F^2 \xi.
 \end{eqnarray}
 
  Theoretically, the correlation function (\ref{eq:3point}) may be written as 
  \begin{eqnarray}
  \label{IWtheo}
   \Xi_{theor}= \int d\nu d\nu' \frac{\rho_{pert}(\nu,\nu')}{(\nu-\omega-i\epsilon)(\nu'-\omega'-i\epsilon)}
    + \Xi_{NP}+subtractions
  \end{eqnarray}
  which can be calculated perturbatively in deep Euclidean region 
($\omega,\omega'\ll \bar{\Lambda}$).
  Note that there are two momentum variables for the correlator (\ref{eq:3point}), 
  a double Borel transformation $\hat{B}^{(\omega')}_{\tau'} \hat{B}^{(\omega)}_{\tau}$ 
  should be applied to both sides of the sum rule. Because of the heavy quark symmetry, 
$\omega$ and $\omega'$ are symmetric in (\ref{eq:3point}), and thus it is natural and 
convenient to choose $\tau=\tau'=2 T$. It is useful to define $\omega_{\pm}=\frac{1}{2} 
(\omega\pm \omega')$, and integrate first the spectral function over $\omega_{-}$ at 
the region $-\nu_{+} <\nu_{-} <\nu_{+}$. Finally the quark-hadron duality allows us to write 
  \begin{eqnarray}
  \label{eq:IWphentheo}
   \tilde{\Xi}_{phen}=2\int^{\omega_0(y)}_{0} d\nu_{+} e^{-\nu_{+}/T} \tilde{\rho}_{pert}(\nu_{+})+\tilde{\Xi}_{NP};
  \end{eqnarray}
 where $\tilde{\Xi}$ denotes the result obtained by applying double Borel operators to $\Xi$, 
and 
  \begin{eqnarray}
    \tilde{\rho}_{pert}(\nu_{+})=\int^{\nu_{+}}_{-\nu_{+}} d\nu_{-} \rho_{pert}(\nu_{+},\nu_{-}).
  \end{eqnarray}
  
 The one loop perturbative diagrams and lowest order nonperturbative diagrams proportional 
to quark condensate and mixed quark-gluon condensate are listed in Fig.2. In this section 
the gluon condensate can be safely neglected since its contribution is 
tiny. Calculation of those Feynman diagrams in Fig.2 gives 
  \begin{eqnarray}
  \label{eq:IWSR}
    && F^2 \xi e^{-\omega_M/T}=\frac{3}{2\pi^2} \int^{\omega_0(y)}_{0} d\omega_{+} e^{-\omega_{+}/T}
       \frac{\omega^2_{+}}{(y+1)^2} \nonumber\\
    && -<\bar{q}q> -i<\bar{q}\sigma_{\alpha\beta} F^{\alpha\beta} q>\frac{1+y}{8T^2}.
  \end{eqnarray}
  The continuum threshold energy in (\ref{eq:IWSR}) is in general a function of the recoil 
variable $y$. One may employ different models for reasonable choice of this function. 
  It is seen that if the $1/m_Q$ order terms and order $\alpha_s$ terms in (\ref{eq:decaySR})
 are neglected, (\ref{eq:IWSR}) reduces to (\ref{eq:decaySR}) at the zero recoil point. 
This implies that we may use the same values of $T$ and $\omega_M$ as those
  in (\ref{eq:decaySR}) for evaluating the wave functions $\xi$, $\kappa_1$ and $\kappa_2$.
The values of $T$ and $\omega_M$ can be read from Fig.4, Fig.5 and Fig.6. 
Note that the threshold energy satisfies the normalization $\omega_0(1)=\omega_0$.
  Our numerical results are plotted in Fig.7, where we have used the values of $\omega_0$, 
  $\omega^{(0)}_{M}$ and $F$ obtained in the previous section. 
 To be consistent, as the QCD radiative corrections are not considered 
 in eq.(\ref{eq:IWSR}), we have used the values given in eq.(\ref{eq:decval1}) where 
 the results were obtained without QCD corrections.
 For comparison, we have used two simple models considered in \cite{neubprd46}:
    \begin{eqnarray}
     \label{eq:model}
       \omega_0(y)=\omega_0(1)
         \left\{
           \begin{array}{cl}
              1, & \mbox{model 1};  \\
              \frac{y+1}{2 y}, & \mbox{model 2}.
           \end{array}
         \right.
     \end{eqnarray}

 We now turn to the calculations for the wave functions $\kappa_i(y)$ defined 
in (\ref{eq:parametrization}). For that, one may consider the following three-point 
correlation function at $1/m_Q$ order:
  \begin{eqnarray}
  \label{eq:3pointkappa}
  {\cal K}(\omega,\omega')&=& \int d^4x d^4y e^{i(P'\cdot x-P\cdot y)}
 <0|T\{(\bar{q} \bar{\Gamma}_{M'} \bar{Q} )_{(x)}, \nonumber \\
  & &  \frac{-1}{2m_Q} \big{(} \bar{Q} \Gamma \frac{P_{+}}{iv\cdot D} 
(i \DSC)^2 Q \big{)}_{(0)}, (\bar{Q} \Gamma_{M} q)_{(y)} \}|0>
\nonumber\\
&=& \int d^4x d^4y e^{i(k'\cdot x-k\cdot y)}
 <0|T\{(\bar{q} \bar{\Gamma}_{M'} \QVP )_{(x)}, \nonumber \\
  &&  \frac{-1}{2m_Q} \big{(} \QVBP \Gamma \frac{P_{+}}{iv\cdot D} 
(i \DSC)^2 \QV \big{)}_{(0)}, (\QVB \Gamma_{M} q)_{(y)} \}|0> +O(1/{m^2_Q})
  \end{eqnarray}
  Saturating this three-point Green function with hadron states, one gets the double-pole 
  contribution 
  \begin{eqnarray}
  \label{eq:kappaphen1}
  &&{\cal K}_{pole}(\omega,\omega') =-(\sum_{pole}) \frac{m_M m_{M'} }
  {m^2_Q \bar{\Lambda}_M \bar{\Lambda}_{M'} (\omega_M-\omega-i\epsilon) 
  (\omega'_M-\omega'-i\epsilon) }\nonumber \\
  && \;\;\; \times <0|\bar{q} \bar{\Gamma}_M \QVP|M'_{v'}><M'_{v'}| \QVBP \Gamma \frac{-1}{2 m_Q}
  \frac{P_{+}}{iv\cdot D} (i \DSC)^2 \QV|M_v>
   <M_v|\QVB \Gamma q|0> .
  \end{eqnarray}

  The heavy-heavy current in (\ref{eq:3pointkappa}) contains both spin-symmetry conserving 
operator $\frac{1}{2m_Q}\QVBP \Gamma \frac{1}{iv\cdot D} (\DC)^2 \QV$ and spin-symmetry 
violating operator $\frac{1}{2m_Q} \QVBP \Gamma \frac{1}{iv\cdot D} P_{+} \frac{i}{2} 
\sigma_{\alpha\beta} F^{\alpha\beta} \QV$. The hadronic matrix elements of the latter 
are parameterized by the wave functions $\kappa_2(y)$ and $\kappa_3(y)$. In this note, 
we may consider only the Feynman diagrams shown in Fig.3, namely we will neglect  
  radiative corrections as a first approximation. In such a treatment the resulting 
contributions from the spin-symmetry violating operator may proportional to the mixed 
quark-gluon condensate. Noticing that in the fixed point gauge
  \begin{eqnarray}
  <0|:\bar{q}(x) A_{\mu}(z) q(0):|0>&=&\frac{z^{\nu}}{96} \sigma_{\nu\mu} 
  <\bar{q}\sigma_{\alpha\beta} F^{\alpha\beta} q > \nonumber\\
  &&\hspace{-2cm} +\mbox{ higher dimensional condensates to be neglected}, 
  \end{eqnarray}
  it is readily seen from (\ref{eq:kappadecomp}) that there would be no contributions to 
$\kappa_3(y)$ at the order we are considering. So in this approximation we have 
$\kappa_3(y)=0$, namely $\kappa_3(y)$ only receives contributions from higher order and 
higher dimensional condensates which are expected to be small. For this reason, 
we may rewrite (\ref{eq:kappaphen1}) as  
  \begin{eqnarray}
  \label{eq:kappaphen}
   {\cal K}_{pole}(\omega,\omega')=-\frac{Tr[\bar{\Gamma}_{M'} P'_{+}\Gamma P_{+} \Gamma_M]}
   {(\omega_M-\omega-i\epsilon)(\omega'_M-\omega'-i\epsilon)} \frac{\bar{\Lambda}^2}
   {\bar{\Lambda}_M \bar{\Lambda}_{M'} }
\frac{m_M m_{M'}}{m^2_Q}
    F^2 \frac{1}{2 m_Q \bar{\Lambda} } (\kappa_1+d_M \kappa_2),
  \end{eqnarray}
 where we have used (\ref{eq:parametrization}) and (\ref{eq:diracsimp}) as well as the formula 
 $P_{+} \sigma_{\mu \nu} {\cal M} (v) \sigma^{\mu \nu}=2 d_M {\cal M}(v)$. In evaluating the 
three-point Green function of (\ref{eq:3pointkappa}), one meets a non-local operator, 
to be convenient of calculating ${\cal K}$, one may choose the axial gauge 
$v\cdot A=0$ \cite{wise}. Note that in this gauge the diagrams with 
 gluons attached to a heavy quark line are absent. 
  
 Adopting the same strategy for evaluating the Isgur-Wise function $\xi(y)$, 
we arrive at the following double Borel transformed sum rule
 \begin{eqnarray}
 \label{eq:kappaSR}
  && -F^2 \frac{1}{2 m_Q \bar{\Lambda} } (\kappa_1+d_M \kappa_2) e^{-\omega_{M}/T}=\frac{1}{2\pi^2}
 \int^{\omega_c}_{0} d\omega_{+} e^{-\omega_{+}/T} \frac{(2y+1)\omega^3_{+}}{m_Q (y+1)^3} \nonumber\\
  &&\hspace{2cm} +i<\bar{q}\sigma_{\alpha\beta} F^{\alpha\beta}q> (\frac{3}{16m_Q T}-\frac{d_M}{24m_Q T}).
  \end{eqnarray}  
 
 By separately considering the spin-conserving and spin-breaking corrections to 
the limit case $m_Q \to \infty$ (or equivalently considering those terms with and without
$d_M$), and taking the central values in (\ref{eq:decval1}), we obtain numerical results 
which are plotted in Fig.8. The curves in Fig.8 are corresponding to the results at 
$T=1.0$GeV which is chosen by considering the stability region of the curves exhibited 
in Fig.4. We also present in Fig.9 amd Fig.10 the values of $\kappa_1(1)$ and $\kappa_2(1)$ 
as functions of the Borel parameter $T$. It is seen from those two figures that $\kappa_1(1)$
 and $\kappa_2(1)$ are really stable at the region around $T=1.0$GeV. The stable regions in 
Fig. 9 and Fig. 10 are consistent each other and also consistent with the ones in Fig.4-Fig.6. 
 With these considerations and analyses, we may present the final result for $\kappa_1(1)$ as
\begin{equation}
\label{eq:kappa1value}
 \kappa_1 \equiv \kappa_1(1)= -0.50 \pm 0.18 \mbox{GeV}^2
\end{equation}
which agrees with the one extracted from the heavy meson masses\cite{wwy1}, 
where $\kappa_1(1)$ could range
from $-0.8 GeV^2$ to $-0.25 GeV^2$ with a favoriable value 
$\kappa_1(1)\approx -0.61 GeV^2$.

  In our previous papers\cite{wwy1,wwy2,wy}, we have argued that $<iv\cdot D>$ is of order 
the binding energy $\bar{\Lambda}$. For simplifying the analyses, we have 
actually made a heavy quark expansion at point $ <iv\cdot D > = \bar{\Lambda}$
for inclusive decays of heavy hadrons\cite{wwy2,wy}. To check the validity of this 
approximation, we may replace the non-local operator $\frac{1}{iv\cdot D}$ in 
 (\ref{eq:parametrization}) with $\frac{1}{\bar{\Lambda}}$ and evaluate the resulting local
matrix element, then a comparison between two results should allow one to test 
the goodness of the approximation. By doing this, we may reparameterize the matrix elements as
 \begin{eqnarray}
 \label{eq:Kpara}
&& <M^{\prime}_{v^{\prime}}\vert \QVBP\Gamma \frac{-1}{\bar{\Lambda}} P_{+} \DC^2 
      \QV \vert M_v> =-K_1(y) \frac{1}{\bar{\Lambda}} Tr[\bar{\cal M}^{\prime}\Gamma {\cal M}],
   \nonumber      \\
&& <M^{\prime}_{v^{\prime}}\vert \QVBP\Gamma \frac{-1}{\bar{\Lambda}} P_{+} \frac{i}{2}
      \sigma_{\alpha\beta}F^{\alpha\beta}\QV \vert M_v> =\frac{1}{\bar{\Lambda}}Tr[K_{\alpha\beta}(v,v^{\prime})
      \bar{\cal M}^{\prime}\Gamma P_{+}\frac{i}{2}\sigma^{\alpha\beta}{\cal M}],
\end{eqnarray} 
 where $K_{\alpha \beta}$ may be decomposed in a similar way as $\kappa_{\alpha\beta}$.
 Applying the sum rule approach once more to evaluate the parameter $K_1$ and following the
 same strategy as before, we yield the following sum rule formula for $K_1$,
 \begin{eqnarray}
  \label{eq:K1sum}
  \frac{1}{2} F^2 K_1 e^{-\omega_M/T} =-\frac{1}{8 \pi^2} \int^{\omega_0}_{0} 
  e^{-\omega_{+}/T} \frac{1+2 y}{(1+y)^3} \omega^4_{+}
  -i <\bar{q}\sigma_{\alpha \beta} F^{\alpha\beta} q> \frac{3}{16},
 \end{eqnarray}
 where the corresponding Feynman diagrams are the same as those in Fig.3 with the box 
 now representing the new operator 
 $-\frac{1}{2m_Q}\QVBP \Gamma \frac{P_{+}}{\bar{\Lambda}} (i\DSC)^2 \QV $. 
 The numerical results of (\ref{eq:K1sum}) are shown in Fig.11 and Fig.12, it is found that
\begin{equation} 
\label{K1value}
 K_1(1) \approx -0.40 \mbox{GeV}^2
\end{equation}
which is slightly lower than $\kappa_1(1)$. Comparing the two nemerical results of $K_1(1)$ 
and $\kappa_1(1)$, we see from (\ref{eq:parametrization}) and  
 (\ref{eq:Kpara}) that the simple replacement
 \begin{eqnarray}
 \label{eq:validity}
 iv\cdot D \sim v\cdot k \approx \bar{\Lambda} 
 \end{eqnarray}
 is actually a reliable approximation for the operator $i v\cdot D$. 

The above demonstration supports the analyses in \cite{wwy1,wy}. We are confirmed to believe 
that the HQEFT is more reliable to describe the off-mass shell heavy quark within heavy 
hadrons and can provide a consistent understanding on both exclusive and inclusion decays 
of heavy hadrons.
 
 We may return to comment on the wave function $\kappa_2(y)$. Up to the order considered above, 
the sum rule formula (\ref{eq:kappaSR}) leads to the value $\kappa_2(1) \approx 0.015$GeV$^2$  
 which is much smaller than the one extracted from the meson masses\cite{wwy1} 
(where $\kappa_2(1)\approx 0.056 \mbox{GeV}^2$). One of possible reasons for such a 
discrepancy may be seen from (\ref{eq:kappaSR}) in which only one mixed condensate term 
contributes to $\kappa_2$. That unique term is relevant to the spin-symmetry breaking term 
and arises from the diagram Fig.3(c). As the operator parameterized by $\kappa_2$ 
in (\ref{eq:parametrization}) contains gluon fields, the lowest order perturbative 
contributions should arise from the two-loop diagram in Fig.11. In order to 
 improve the determination for $\kappa_2$, one should therefore calculate at least 
two-loop perturbative diagrams as well as nonperturbative diagrams up to order $\alpha_s$  
in a consistent way. Since all these diagrams are at least at the order of $\alpha_s$, 
the more precise value of $\kappa_2$ should not be too large. We should not present 
calculations of the additional diagrams for $\kappa_2$ in this paper.
 
 In fact there is another way to estimate the values of $\kappa_1(1)$ and $\kappa_2(1)$ 
directly from the results obtained through two-point correlator functions in Sec.\ref{decay}. 
This is again because the remarkable relation between the heavy-light hadron mass and wave functions 
at zero recoil resulted from the complete HQEFT. It can be seen from Eqs.(\ref{massrelation}) and 
(\ref{eq:defomegacor}) that the form factors $\kappa_1(1)$ and $\kappa_2(1)$ are simply given by
\begin{eqnarray} 
 \kappa_1(1)=-\frac{\omega^{(1)}_M}{2}, \; \; \; \kappa_2(1)=-\frac{\omega^{(2)}_M}{2}.
\end{eqnarray}  
With these relations, we obtain from the values of $\omega^{(i)}_M$ in eq.(\ref{eq:decval2}) 
\begin{eqnarray} 
 \kappa_1(1) \approx -0.43 \mbox{GeV}^2, \; \; \; \kappa_2(1) \approx 0.08 \mbox{GeV}^2,
\end{eqnarray}  
which are consistent with the results yielded above (eq.(\ref{eq:kappa1value})) and also those 
obtained in Ref.\cite{wwy1} for $\kappa_1(1)$ and $\kappa_2(1)$.

 \section{Coorections From Two-loop Perturbative QCD}\label{radiative}
 
 In order to take a look at the magnitude of the effects of QCD radiative corrections, 
as in Ref.\cite{neu1076}, one can now include the two-loop perturbative contributions. 
Their effects can be simply taken into account by replacing the perturbative contributions 
in the sum rule (\ref{eq:beforeBT}) with the following ones 
 \begin{eqnarray}
 \label{eq:beforeBT2loop}
  \Pi_{pert}(\omega)=
   \frac{3}{8\pi^2} \int^{\omega_c}_{0} d \nu 
   (\frac{1}{4 m_Q}+\frac{1}{\nu-\omega-i\epsilon} )
   \nu^2 (1+\frac{2 \alpha_s}{\pi} [ ln\frac{2 \bar{\Lambda}}{\nu} +\frac{13}{6}
   +\frac{2 \pi^2}{9} ] -\frac{3 \nu }{4 m_Q}).
 \end{eqnarray} 
In comparison with the results from leading QCD corrections, we have plotted the modified results 
with two-loop QCD corrections in Figs. (14-22). As a consequence, 
instead of eqs.(\ref{eq:decval1}) and (\ref{eq:decval2}), we arrive at the following 
modified results
 \begin{eqnarray}
  \label{eq:decval12loop}
   \omega_0&=&1.8\pm 0.3 \; \mbox{GeV},\nonumber\\
   \frac{\omega^{(0)}_M}{2}&=&\bar{\Lambda}=0.56\pm 0.08 \; \mbox{GeV},\nonumber\\
   F&=&0.38\pm 0.06 \; \mbox{GeV}^{3/2}.
  \end{eqnarray}
and
   \begin{eqnarray}
  \label{eq:decval22loop}
   \omega_1 &=& 1.0 \pm 0.2 \; \mbox{GeV}^2, \nonumber\\
   \omega^{(1)}_M &=& 0.65 \pm 0.10 \; \mbox{GeV}^2, \nonumber\\
   G_1&=&0.75 \pm 0.15 \; \mbox{GeV}, \nonumber \\
   g_1&=&0.46 \pm 0.12 \; \mbox{GeV}, \nonumber \\
   \omega_2 &=& -0.15\pm 0.05 \;  \mbox{GeV}^2, \nonumber\\
   \omega^{(2)}_M &=& -0.14\pm 0.03 \; \mbox{GeV}^2, \nonumber\\
   G_2 &=& -0.09\pm 0.03 \; \mbox{GeV}, \nonumber \\
   g_2 &=& -0.06\pm 0.02 \; \mbox{GeV}.
  \end{eqnarray}
 Correspondingly, instead of eq.(\ref{eq:decval3QCD}), we have 
\begin{eqnarray}
  \label{eq:decval3QCD2loop}
   &f_B(m_b)=0.196 \pm 0.044  \; \mbox{GeV}, \hspace{2cm}
   &f_{B^{\ast}}(m_b)=0.206 \pm 0.039 \;  \mbox{GeV},\nonumber\\
   &f_D(m_c)=0.298 \pm 0.109  \; \mbox{GeV}, \hspace{2cm}
   &f_{D^{\ast}}(m_c)=0.354 \pm 0.090 \;  \mbox{GeV}.
 \end{eqnarray}
 Comparing these values with those obtained in Sec.\ref{decay}, we see that the QCD 
 radiative corrections may enlarge $F$ by about 25$\%$. It implies that the radiative 
 corrections may be significant for a more accurate determination for some physical quantities.
The large values seems to be consistent with the ones from the recent calculations by
Lattice QCD approach\cite{LQCD3,LQCD4,LQCD5}.
 
 With those values in eq.(\ref{eq:decval12loop}), we yield from the sum rules in 
 eqs.(\ref{eq:kappaSR}) and (\ref{eq:K1sum})
\begin{eqnarray}
\label{eq:kappa1value2loop}
\kappa_1(1)&=&-0.34 \mbox{GeV}^2;  \\
\label{eq:K1value2loop} 
 K_1(1)&=&-0.26   \mbox{GeV}^2.
\end{eqnarray}
 which are lower than the ones without two-loop perturbative QCD corrections. While we would like to 
point out that the modified results for $\kappa_1(1)$ and $K_1(1)$  
( eqs.(\ref{eq:kappa1value2loop}) and (\ref{eq:K1value2loop}) ) may not be regarded to be more reliable 
than the ones given in eqs.(\ref{eq:kappa1value}) and (\ref{K1value}) since eqs.(\ref{eq:kappaSR}) 
and (\ref{eq:K1sum}) contain only the leading order contributions in perturbation theory. For a complete and 
consistent evaluation, one should also include the next-to-leading order corrections to  
eqs.(\ref{eq:kappaSR}) and (\ref{eq:K1sum}). 
 But comparing the two values in eqs.(\ref{eq:kappa1value2loop}) and (\ref{eq:K1value2loop}) 
 we see that the approxmation $iv\cdot D \sim v\cdot k \simeq \bar{\Lambda}$ still holds, 
 though both the values of $\kappa_1(1)$ and $K_1(1)$ are now smaller than those presented in 
 Sec.\ref{tran} because the input parameter $F$ is enlarged by the two-loop perturbative 
 corrections. 
 
 It is also interesting to notice that the relation $\kappa_1(1)=-\omega^{(1)}_M/2\simeq -0.37\mbox{GeV}^2$ 
 is still satisfied well by looking at the result given in eq.(\ref{eq:kappa1value2loop}) and 
the value of $\omega^{(1)}_M$ in eq.(\ref{eq:decval22loop}). In conclusion, all the numerical 
results in this paper turn out to be consistent.
 
 \section{Conclusions and Remarks}\label{sum}

 We have present, within the complete HQEFT, a consistent evaluation for the decay constants 
and wave functions of heavy-light hadron systems up to order of $1/m_Q$. It has been seen that
the QCD Lagrangian for heavy quarks do neet to be transformed into a new heavy quark effective 
Lagrangian with including contributions of both particle fields and antiparticle fields\cite{ylw},
and the currents containing heavy quark can be consistently expanded in powers of $1/m_Q$. 
Though the leading operator in the $1/m_Q$ expansion is (must be) the same as the one in the 
usual HQET, the operators at order $1/m_Q$ begin to be different from those in the usual HQET. 
 
 In the complete HQEFT, corrections arising from both the current expansion and the insertion 
 of Lagrangian into the heavy-light meson anahilation matrix elements are characterized by 
two factors $G_1$ and $G_2$. Similarly, corrections arising from both sources to the heavy
 meson transition matrix elements are characterized by three functions $\kappa_1(y)$, $\kappa_2(y)$
 and $\kappa_3(y)$. Furthermore, the values of $\kappa_1$ and $\kappa_2$ at the zero recoil point 
also characterize the $1/m_Q$ order corrections to the meson masses. These makes the HQEFT be 
much elegant than the usual HQET. 
 
 The leading order contributions and the $1/m_Q$ corrections to heavy meson decay constants and 
 heavy meson transition matrix elements have been investigated consistently by using QCD sum rule approach 
within the framework of HQEFT of QCD. 
For heavy-light meson decays, we have calculated the form factor $F$ at leading order 
($m_Q\rightarrow\infty$), and the form factors $G_1$ and $G_2$ concerned at the $1/m_Q$ order as well as 
 the binding energy $\bar{\Lambda}$. Particularly, we have found that the $1/m_Q$ 
 corrections to the heavy-light meson decay constants are actually determined by two composite
form factors $g_1$ and $g_2$. These two composited form factors were found to be much smaller than 
the heavy quark masses, which implies that the scaling law of the decay constants are only sligtly 
breakdown. This observation  shows that the $1/m_Q$ expansion 
in the conplete HQEFT works well, which is unlike the usual HQET which may lead to, 
as shown in \cite{neu1076}, the breakdown of the $1/m_Q$ expansion in evaluating the decay constants. 
Our results for the heavy-light meson decay constants have also shown a good agreement 
with the known experimental results and upper limits. 
 
 We have also calculated the Isgur-Wise function and $1/m_Q$ order spin-symmetry conserving 
 form factor $\kappa_1$ as functions of the recoil value. 
 It have been found that $\kappa_1(1)\approx -0.50 \pm 0.18 \mbox{GeV}^2$, which agrees with
 the value extracted from the interesting relations between meson mass and wave functions at zero 
 recoil. We have also illustrated how the simple replacement 
 $iv\cdot D \sim v\cdot k \approx \bar{\Lambda}$ holds. This shows that the residual momentum of 
heavy quark within heavy-light hadron does be around the binding energy which had been seen\cite{wwy2} to be 
 the main point to understand the puzzle of the bottom hadron lifetime differences. 
 
 The spin symmetry breaking factor $\kappa_2$ has been discussed. The only diagram at leading order 
yields a $\kappa_2$ value which is much smaller than that obtained in \cite{wwy1}.
Thought its value has not yet been evaluated accuratly, it implies that  $\kappa_2$ must be small as it 
characterizes the spin symmetry breaking effects of heavy-light mesons. A further calculation of 
$\kappa_2$ including higher order contributions remains an interesting subject. 
$\kappa_3(y)$ was also found to be small since it receives contributions only from higher 
order radiative diagrams and higher dimensional condensates. 

 An interesting feature resulting from the HQEFT is that the values of $\kappa_1(1)$ and $\kappa_2(1)$ 
 can also be simply obtained, due to the interesting relation between heavy-light meson mass and wave 
functions at zero recoil, from two-point Green's function in evaluating the heavy-light meson decay 
constants via sum rule approach. It is remarkable that the resulting values of $\kappa_1(1)$ and 
$\kappa_2(1)$ in this way do agree with the results obtained from the analysis of the three-point 
Green's function. Finally, we have shown that the higher order radiative corrections could be nontrival 
for a more accurate calculation of heavy-light meson decays. But the relations
 $iv\cdot D\sim iv\cdot k \simeq  \bar{\Lambda}$ and $\kappa_1(1)=-\omega^{(1)}_M/2$, 
$\kappa_2(1)=-\omega^{(2)}_M/2$ hold even when higher radiative corrections are included. 

 In summary, the complete HQEFT works well for describing the slightly off-mass shell heavy quark within 
heavy hadrons. In this paper, we have further checked the consistent of the HQEFT in applying to the 
heavy-light hadron systems with $m_Q >> \bar{\Lambda} >> m_q$.  For heavy-heavy bound state systems, such as 
bottom-charm system like $B_c$, and muonium $(\mu e)$ system in QED, one needs to make $1/M$ expansion 
for both heavy particles and to redefine the effective fields and bound states when applying for 
the complete heavy particle effective field theory with keeping the antiparticle contributions\cite{ylw}.
 
\acknowledgments

We would like to thank  D. Chang, H.Y. Cheng, Y.B. Dai, H.N. Li and W.M. Zhang for 
useful discussions. This work was supported in part by the NSF of China under the grant No. 19625514.



\newpage
\centerline{\large{FIGURES}}

\begin{center}
\begin{picture}(400,400)(0,0)

\SetWidth{2}
\CArc(50,360)(25,0,180)
\SetWidth{0.5}
\CArc(50,360)(25,180,0)
\DashLine(10,360)(25,360){3}
\DashLine(75,360)(90,360){3}

\SetWidth{2}
\CArc(150,360)(25,0,180)
\SetWidth{0.5}
\CArc(150,360)(25,180,240)
\CArc(150,360)(25,300,360)
\Vertex(140,338){3}
\Vertex(160,338){3}
\DashLine(110,360)(125,360){3}
\DashLine(175,360)(190,360){3}

\SetWidth{2}
\CArc(250,360)(25,0,180)
\SetWidth{0.5}
\CArc(250,360)(25,180,240)
\CArc(250,360)(25,300,360)
\Vertex(240,338){3}
\Vertex(260,338){3}
\Gluon(250,385)(250,360){3}{2}
\Vertex(250,360){3}
\DashLine(210,360)(225,360){3}
\DashLine(275,360)(290,360){3}

\SetWidth{2}
\CArc(350,360)(25,0,180)
\SetWidth{0.5}
\CArc(350,360)(25,180,240)
\CArc(350,360)(25,300,360)
\Vertex(340,338){3}
\Vertex(360,338){3}
\GlueArc(350,390)(20,220,320){3}{2}
\DashLine(310,360)(325,360){3}
\DashLine(375,360)(390,360){3}


\SetWidth{2}
\CArc(50,260)(25,0,180)
\SetWidth{0.5}
\CArc(50,260)(25,180,240)
\CArc(50,260)(25,300,360)
\Vertex(40,238){3}
\Vertex(60,238){3}
\GlueArc(15,280)(35,-70,10){3}{3}
\DashLine(10,260)(25,260){3}
\DashLine(75,260)(90,260){3}

\SetWidth{2}
\CArc(150,260)(25,0,180)
\SetWidth{0.5}
\CArc(150,260)(25,180,240)
\CArc(150,260)(25,300,360)
\Vertex(140,238){3}
\Vertex(160,238){3}
\GlueArc(185,280)(35,170,250){3}{3}
\DashLine(110,260)(125,260){3}
\DashLine(175,260)(190,260){3}

\SetWidth{2}
\CArc(250,260)(25,0,180)
\SetWidth{0.5}
\CArc(250,260)(25,180,240)
\CArc(250,260)(25,300,360)
\Vertex(240,238){3}
\Vertex(260,238){3}
\Gluon(230,250)(270,250){-3}{3}
\DashLine(210,260)(225,260){3}
\DashLine(275,260)(290,260){3}

\SetWidth{2}
\CArc(350,260)(25,0,180)
\SetWidth{0.5}
\CArc(350,260)(25,180,0)
\GlueArc(350,290)(25,220,260){3}{1}
\GlueArc(350,290)(25,280,320){3}{1}
\Vertex(355,265){3}
\Vertex(345,265){3}
\DashLine(310,260)(325,260){3}
\DashLine(375,260)(390,260){3}


\SetWidth{2}
\CArc(50,160)(25,0,180)
\SetWidth{0.5}
\CArc(50,160)(25,180,0)
\GlueArc(50,130)(25,40,80){3}{1}
\GlueArc(50,130)(25,100,140){3}{1}
\Vertex(55,155){3}
\Vertex(45,155){3}
\DashLine(10,160)(25,160){3}
\DashLine(75,160)(90,160){3}

\SetWidth{2}
\CArc(150,160)(25,0,180)
\SetWidth{0.5}
\CArc(150,160)(25,180,0)
\Gluon(150,185)(150,170){3}{1}
\Gluon(150,150)(150,135){3}{1}
\Vertex(150,170){3}
\Vertex(150,150){3}
\DashLine(110,160)(125,160){3}
\DashLine(175,160)(190,160){3}

\SetWidth{2}
\CArc(250,160)(25,0,180)
\SetWidth{0.5}
\CArc(250,160)(25,180,0)
\Gluon(250,185)(250,170){3}{1}
\Gluon(250,200)(250,185){3}{1}
\Vertex(250,170){3}
\Vertex(250,200){3}
\DashLine(210,160)(225,160){3}
\DashLine(275,160)(290,160){3}

\SetWidth{2}
\CArc(350,160)(25,0,180)
\SetWidth{0.5}
\CArc(350,160)(25,180,240)
\CArc(350,160)(25,300,360)
\Vertex(340,138){3}
\Vertex(360,138){3}
\GlueArc(340,138)(12,20,140){3}{1}
\Vertex(350,145){3}
\DashLine(310,160)(325,160){3}
\DashLine(375,160)(390,160){3}


\SetWidth{2}
\CArc(50,60)(25,0,180)
\SetWidth{0.5}
\CArc(50,60)(25,180,240)
\CArc(50,60)(25,300,360)
\Vertex(40,38){3}
\Vertex(60,38){3}
\GlueArc(60,38)(12,40,160){3}{1}
\Vertex(50,45){3}
\DashLine(10,60)(25,60){3}
\DashLine(75,60)(90,60){3}


\end{picture}
\end{center}

Fig.1. Feynman diagrams contributing to heavy meson decay. 
The thick lines are heavy quarks; the light lines are light quarks; 
the curves are gluon fields; the black dots represent condensates; 
and the external dashed lines are the heavy-light currents considered 
in (\ref{eq:2point}). 

\small
\begin{center}
\begin{picture}(400,250)(0,0)
\SetWidth{2}
\Line(20,150)(50,200)
\Line(50,200)(80,150)
\Photon(50,220)(50,200){3}{2}
\SetWidth{0.5}
\Line(20,150)(80,150)
\DashLine(0,150)(20,150){3}
\DashLine(80,150)(100,150){3}

\SetWidth{2}
\Line(170,150)(200,200)
\Line(200,200)(230,150)
\Photon(200,220)(200,200){3}{2}
\SetWidth{0.5}
\Line(170,150)(190,150) \Vertex(190,150){3}
\Line(210,150)(230,150) \Vertex(210,150){3}
\DashLine(150,150)(170,150){3}
\DashLine(230,150)(250,150){3}

\SetWidth{2}
\Line(320,150)(350,200)
\Line(350,200)(380,150)
\Photon(350,220)(350,200){3}{2}
\SetWidth{0.5}
\Line(320,150)(340,150) \Vertex(340,150){3}
\Line(360,150)(380,150) \Vertex(360,150){3}
\GlueArc(343,150)(10,40,180){3}{1}
\Vertex(350,160){3}
\DashLine(300,150)(320,150){3}
\DashLine(380,150)(400,150){3}

\SetWidth{2}
\Line(20,50)(50,100)
\Line(50,100)(80,50)
\Photon(50,120)(50,100){3}{2}
\SetWidth{0.5}
\Line(20,50)(40,50) \Vertex(80,50){3}
\Line(60,50)(80,50) \Vertex(60,50){3}
\GlueArc(57,50)(10,0,140){3}{1}
\Vertex(50,60){3}
\DashLine(0,50)(20,50){3}
\DashLine(80,50)(100,50){3}

\SetWidth{2}
\Line(170,50)(200,100)
\Line(200,100)(230,50)
\Photon(200,120)(200,100){3}{2}
\SetWidth{0.5}
\Line(170,50)(190,50) \Vertex(190,50){3}
\Line(210,50)(230,50) \Vertex(210,50){3}
\Gluon(185,75)(200,60){3}{1}
\Vertex(200,60){3}
\DashLine(150,50)(170,50){3}
\DashLine(230,50)(250,50){3}

\SetWidth{2}
\Line(320,50)(350,100)
\Line(350,100)(380,50)
\Photon(350,120)(350,100){3}{2}
\SetWidth{0.5}
\Line(320,50)(340,50) \Vertex(340,50){3}
\Line(360,50)(380,50) \Vertex(360,50){3}
\Gluon(350,60)(365,75){3}{1}
\Vertex(350,60){3}
\DashLine(300,50)(320,50){3}
\DashLine(380,50)(400,50){3}


\end{picture}
\end{center}

Fig.2. The lowest order Feynman diagrams contributing to $\xi$. 
The wave lines represent the heavy-heavy current $\QVBP \Gamma \QV$ in (\ref{eq:3point}). 

\small
\begin{center}
\begin{picture}(400,250)(0,0)
\SetWidth{2}
\Line(20,150)(50,200)
\Line(50,200)(80,150)
\Photon(50,220)(50,200){3}{2}
\SetWidth{0.5}
\Line(20,150)(80,150)
\BBoxc(50,200)(8,8)
\DashLine(0,150)(20,150){3}
\DashLine(80,150)(100,150){3}

\SetWidth{2}
\Line(170,150)(200,200)
\Line(200,200)(230,150)
\Photon(200,220)(200,200){3}{2}
\SetWidth{0.5}
\Line(170,150)(190,150) \Vertex(190,150){3}
\Line(210,150)(230,150) \Vertex(210,150){3}
\BBoxc(200,200)(8,8)
\DashLine(150,150)(170,150){3}
\DashLine(230,150)(250,150){3}

\SetWidth{2}
\Line(320,150)(350,200)
\Line(350,200)(380,150)
\Photon(350,220)(350,200){3}{2}
\SetWidth{0.5}
\Line(320,150)(340,150) \Vertex(340,150){3}
\Line(360,150)(380,150) \Vertex(360,150){3}
\GlueArc(343,150)(10,40,180){3}{1}
\Vertex(350,160){3}
\BBoxc(350,200)(8,8)
\DashLine(300,150)(320,150){3}
\DashLine(380,150)(400,150){3}

\SetWidth{2}
\Line(20,50)(50,100)
\Line(50,100)(80,50)
\Photon(50,120)(50,100){3}{2}
\SetWidth{0.5}
\Line(20,50)(40,50) \Vertex(80,50){3}
\Line(60,50)(80,50) \Vertex(60,50){3}
\GlueArc(57,50)(10,0,140){3}{1}
\Vertex(50,60){3}
\BBoxc(50,100)(8,8)
\DashLine(0,50)(20,50){3}
\DashLine(80,50)(100,50){3}

\SetWidth{2}
\Line(170,50)(200,100)
\Line(200,100)(230,50)
\Photon(200,120)(200,100){3}{2}
\SetWidth{0.5}
\Line(170,50)(190,50) \Vertex(190,50){3}
\Line(210,50)(230,50) \Vertex(210,50){3}
\Gluon(200,60)(200,100){3}{3}
\Vertex(200,60){3}
\BBoxc(200,100)(8,8)
\DashLine(150,50)(170,50){3}
\DashLine(230,50)(250,50){3}

\end{picture}
\end{center}

Fig.3. The lowest order Feynman diagrams contributing to $\kappa_1$ and $\kappa_2$. 
The box at the up of each diagram represent the $1/m_Q$ order heavy-heavy current 
$-\frac{1}{2m_Q}\QVBP \Gamma \frac{P_{+}}{iv\cdot D} (i\DSC)^2 \QV $ in (\ref{eq:3pointkappa}). 

\newcommand{\PIC}[2]
{
\begin{center}
\begin{picture}(300,300)(0,0)
\put(40,25){
\epsfxsize=8cm
\epsfysize=8cm
\epsffile{#1} }
\put(150,40){\makebox(0,0){#2}}
\end{picture}
\end{center}
}

\newpage
\small
\mbox{}

\PIC{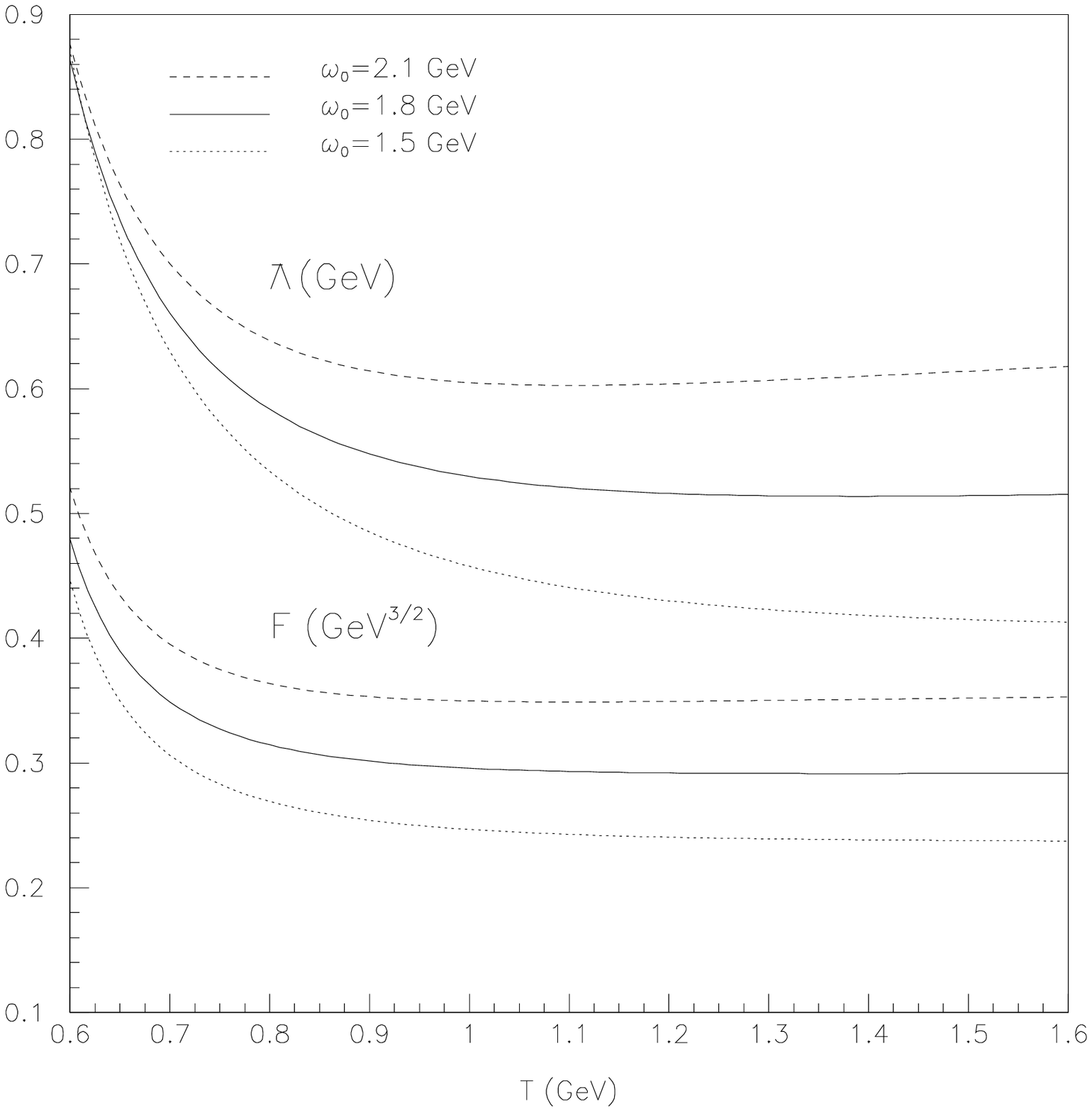}{ Fig.4. Leading oreder sum rule result for heavy meson decay. }

\PIC{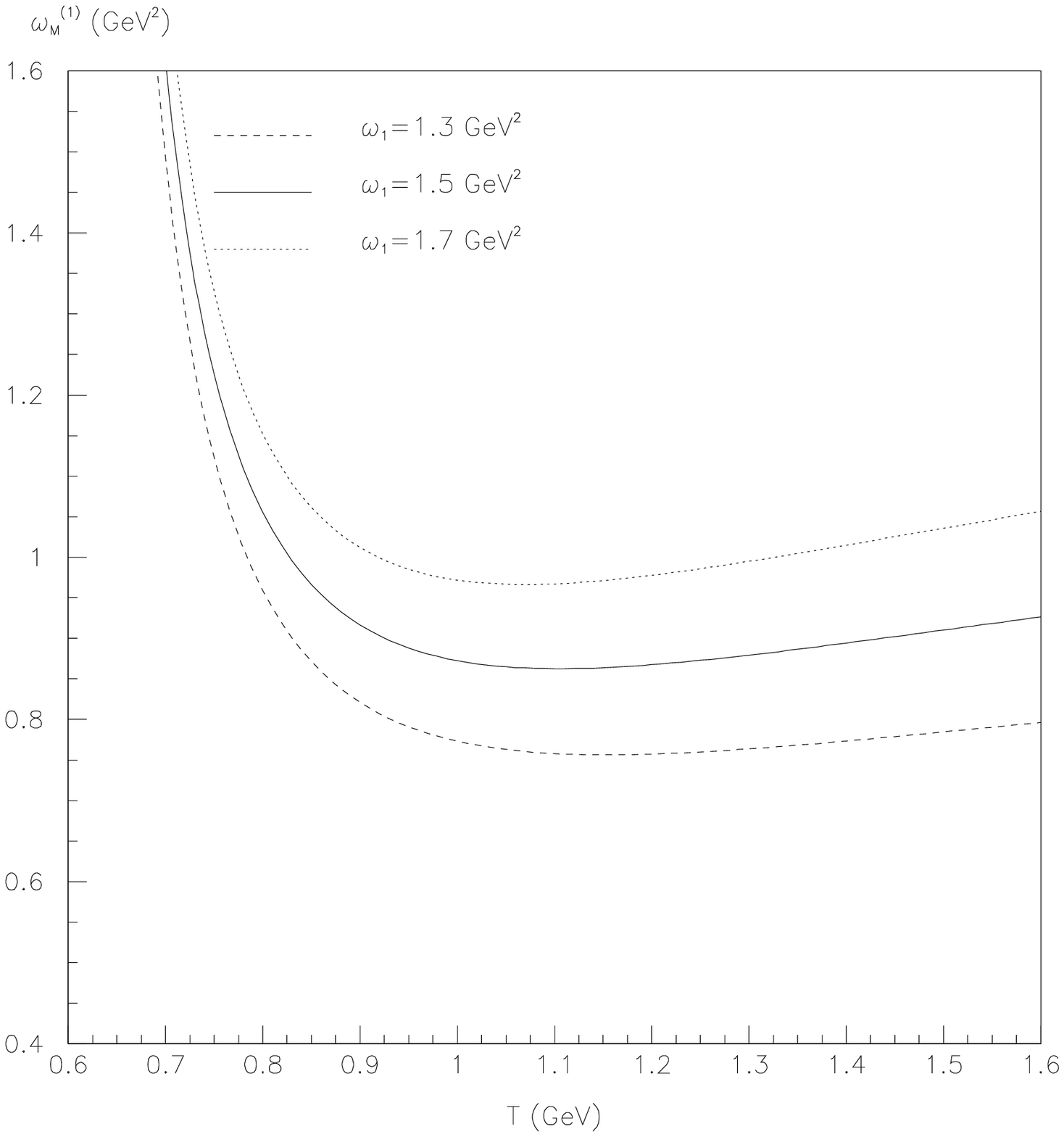}{Fig.5a}

\PIC{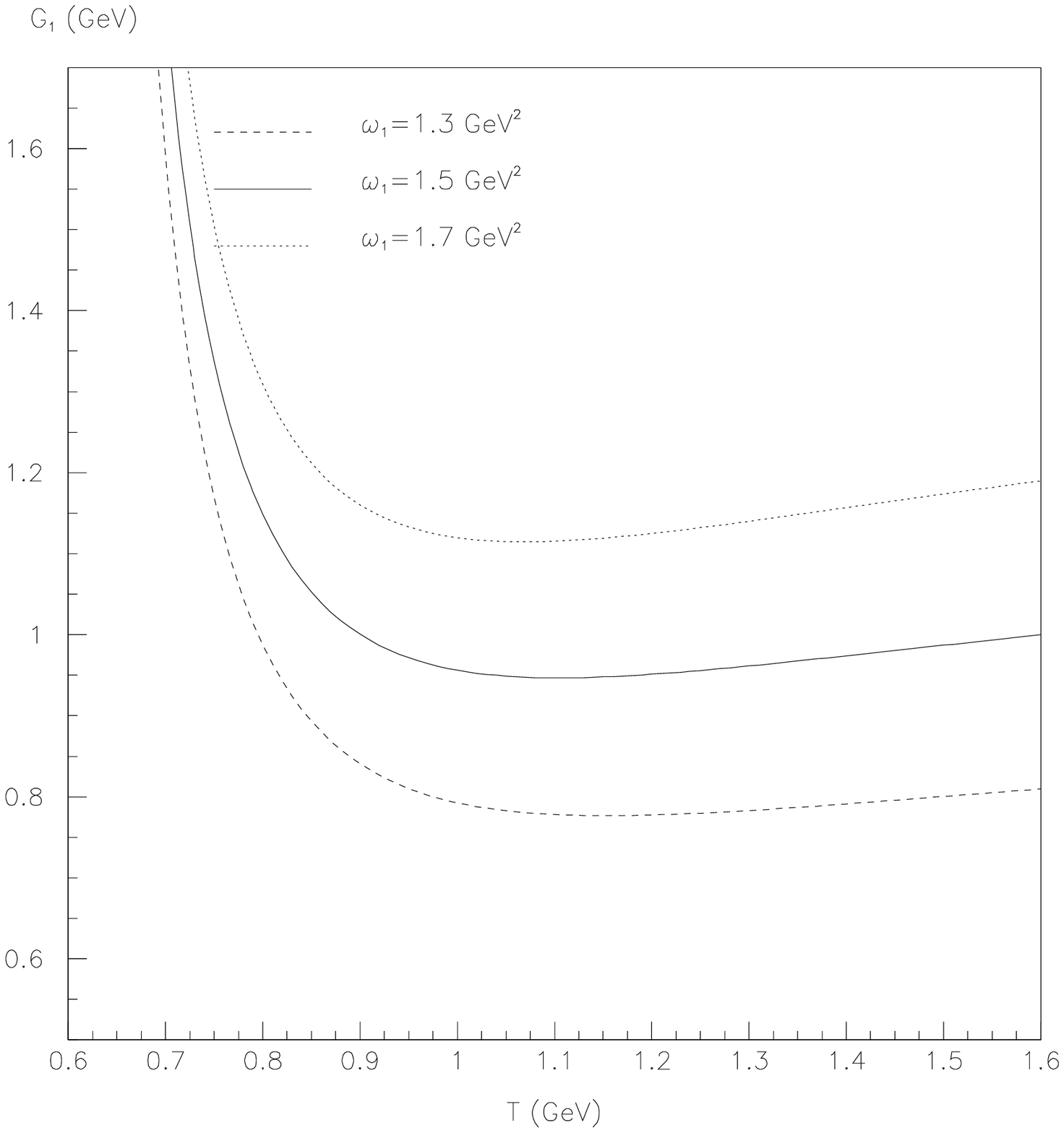}{Fig.5b}

\PIC{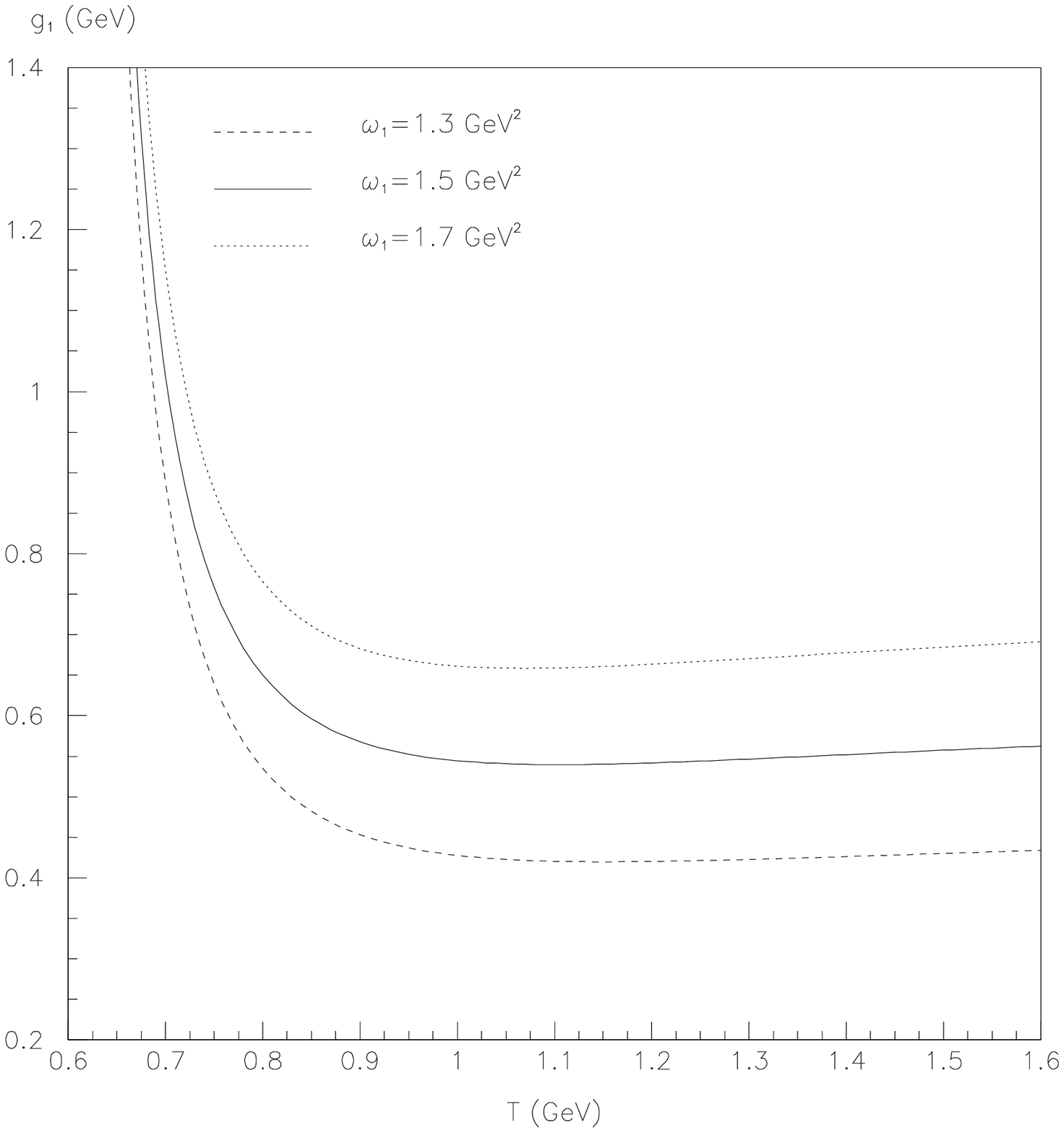}{Fig.5c}  

\vspace{-2cm}{\center{Fig.5 Sum rule result for $1/m_Q$ order spin-symmetry 
conserving corrections to heavy meson decay.}
}

\PIC{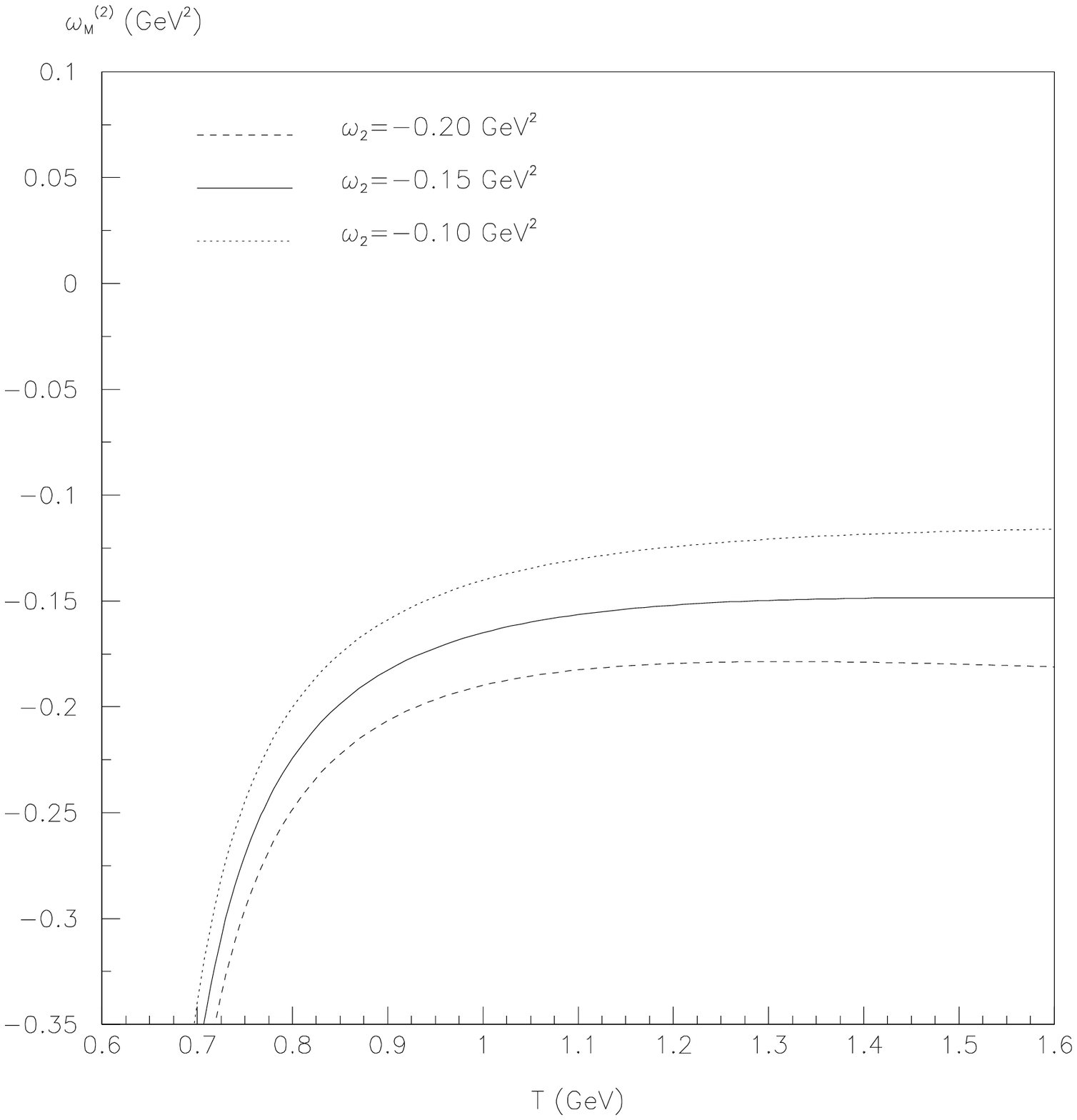}{Fig.6a}

\PIC{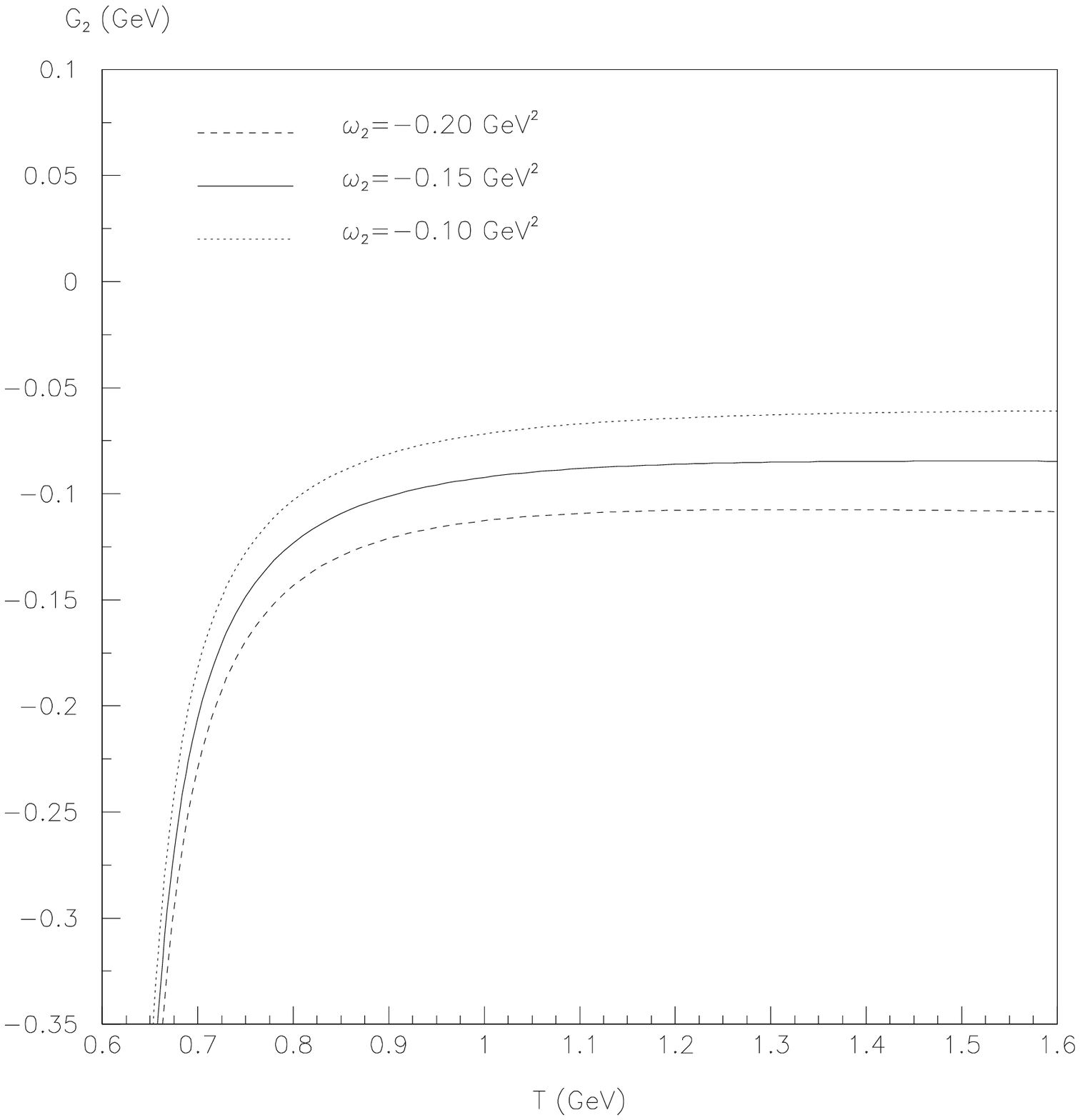}{Fig.6b}

\PIC{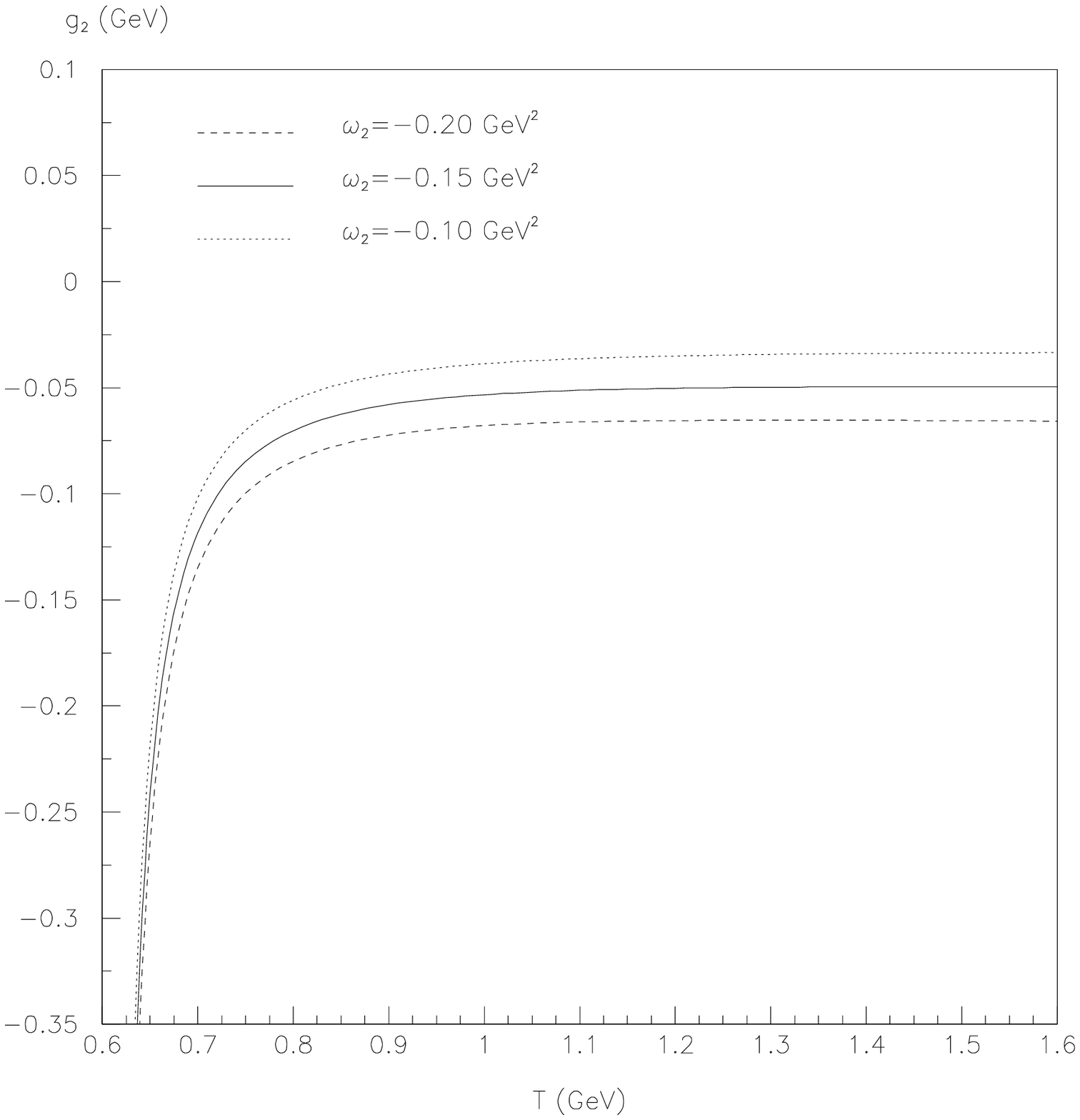}{Fig.6c}

\vspace{-2cm}{\center{Fig.6. Sum rule result for $1/m_Q$ order spin-symmetry 
breaking corrections to heavy meson decay.}
}

\vspace{2cm}

\PIC{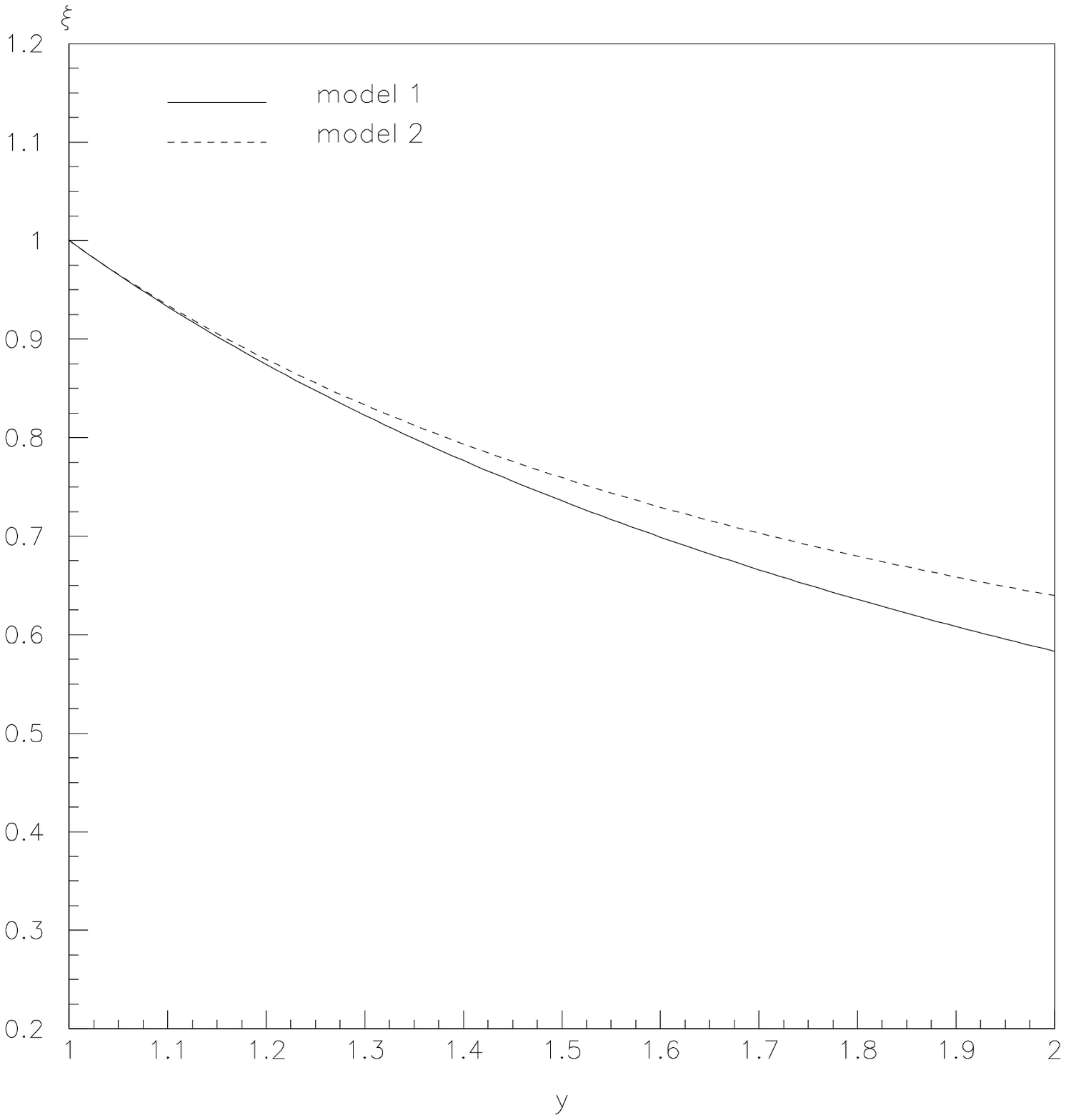}{ Fig.7. Sum rule result for Isgur-Wise function $\xi(y)$. }

\PIC{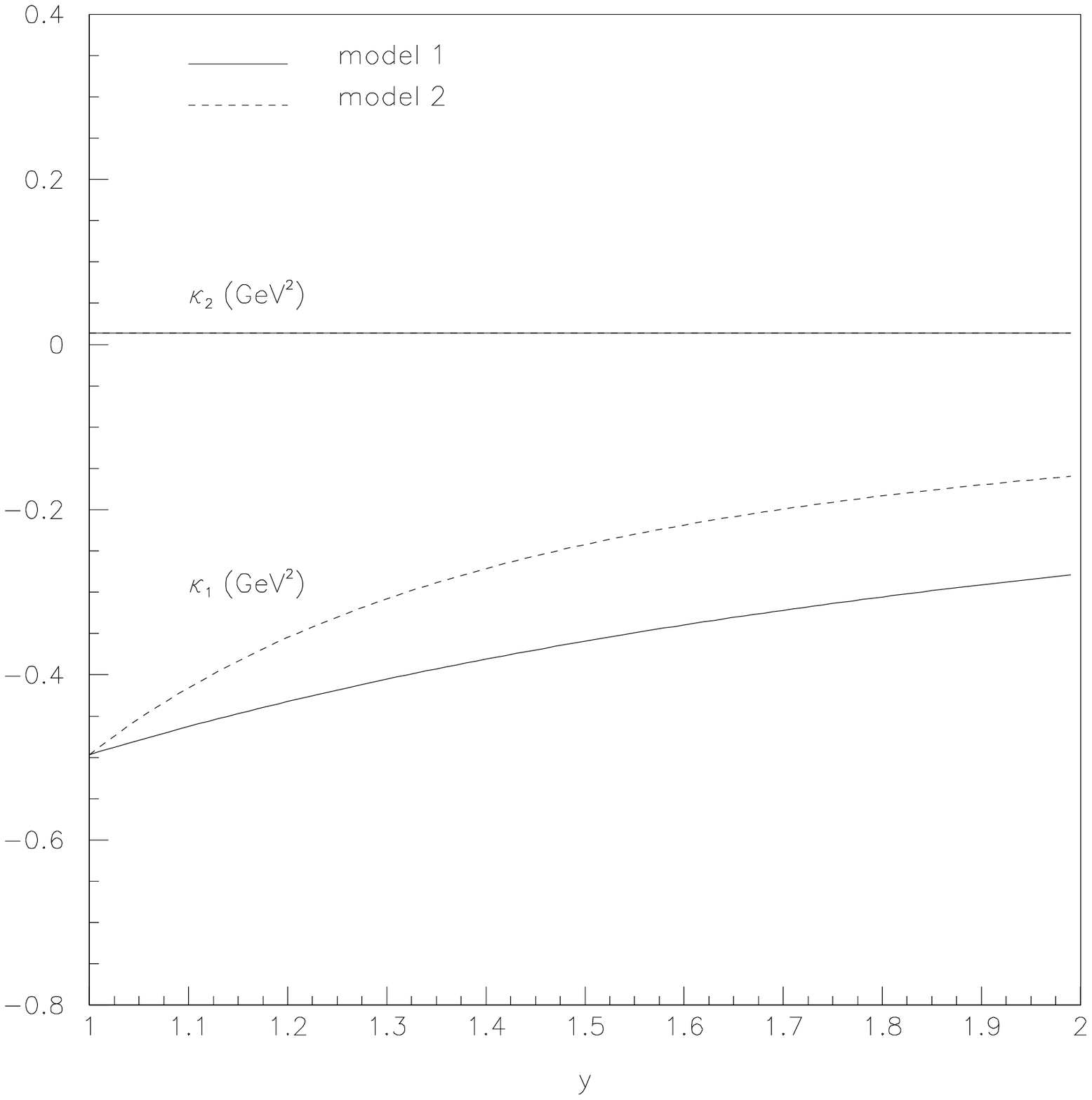}{ Fig.8. Sum rule result for $1/m_Q$ order transition form 
factor $\kappa_1$ and $\kappa_2$. }

\PIC{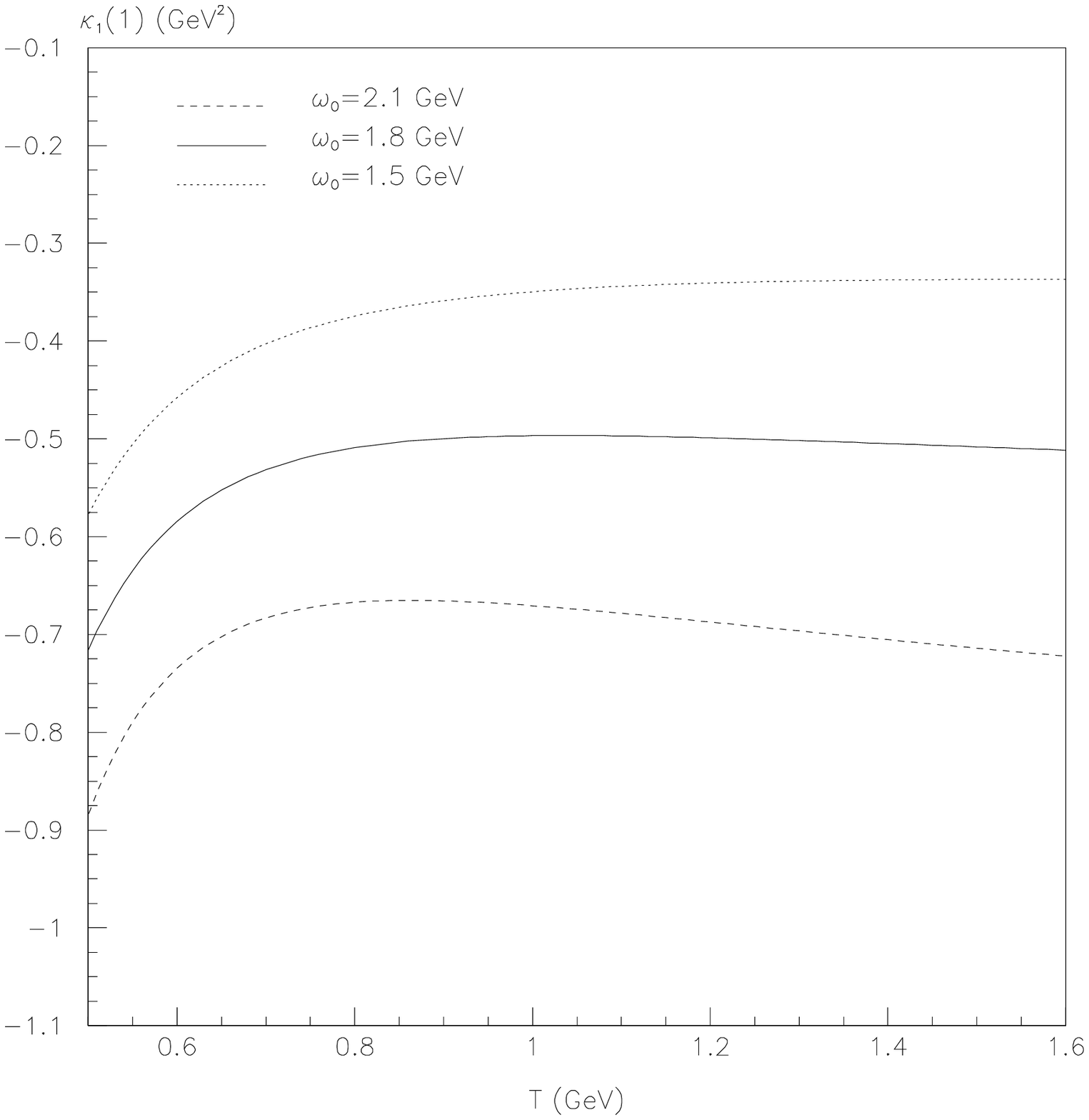}{ Fig.9. $\kappa_1(1)$ as a function of Borel parameter $T$. }

\PIC{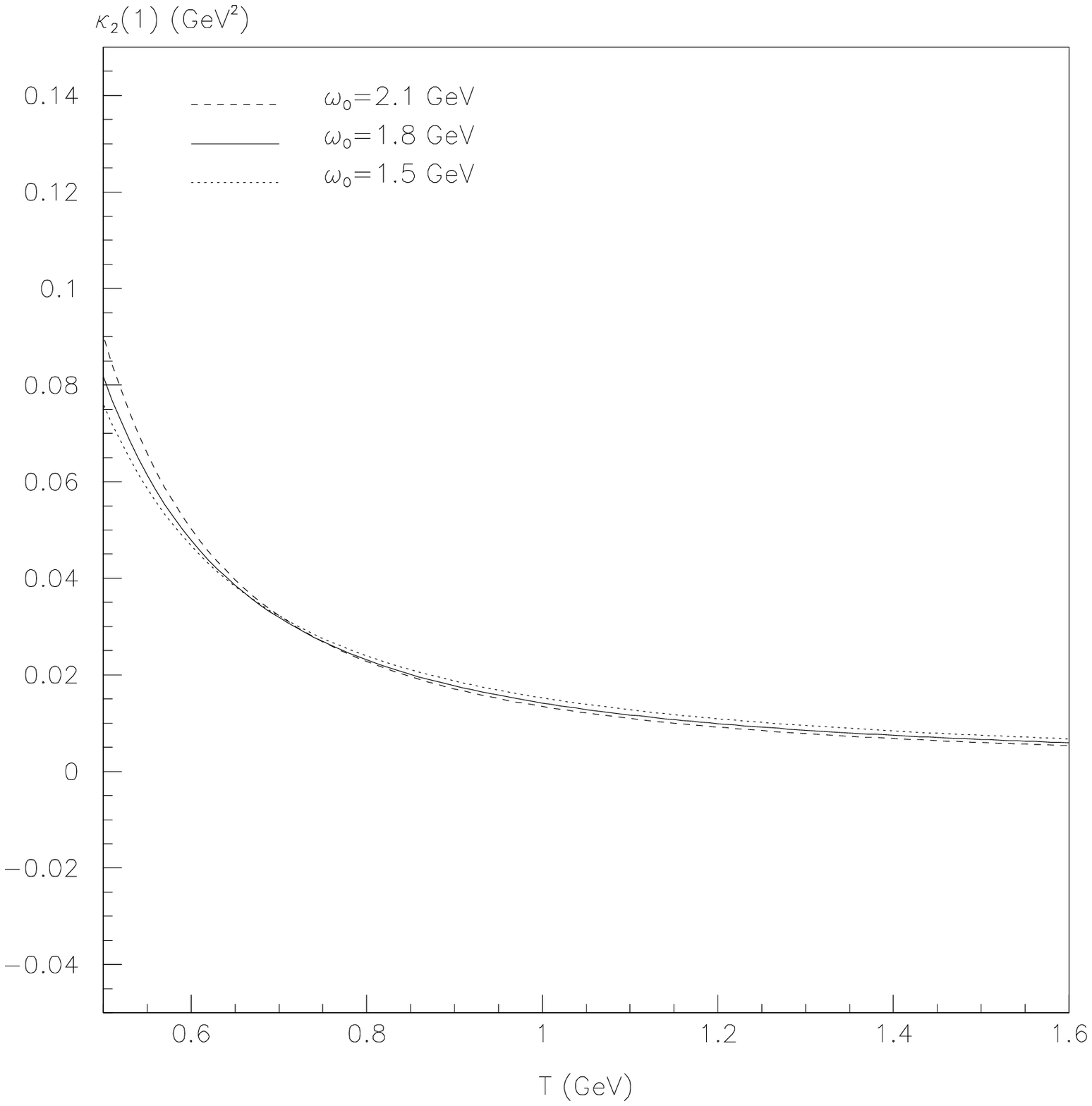}{ Fig.10. $\kappa_2(1)$ as a function of Borel parameter $T$. }

\PIC{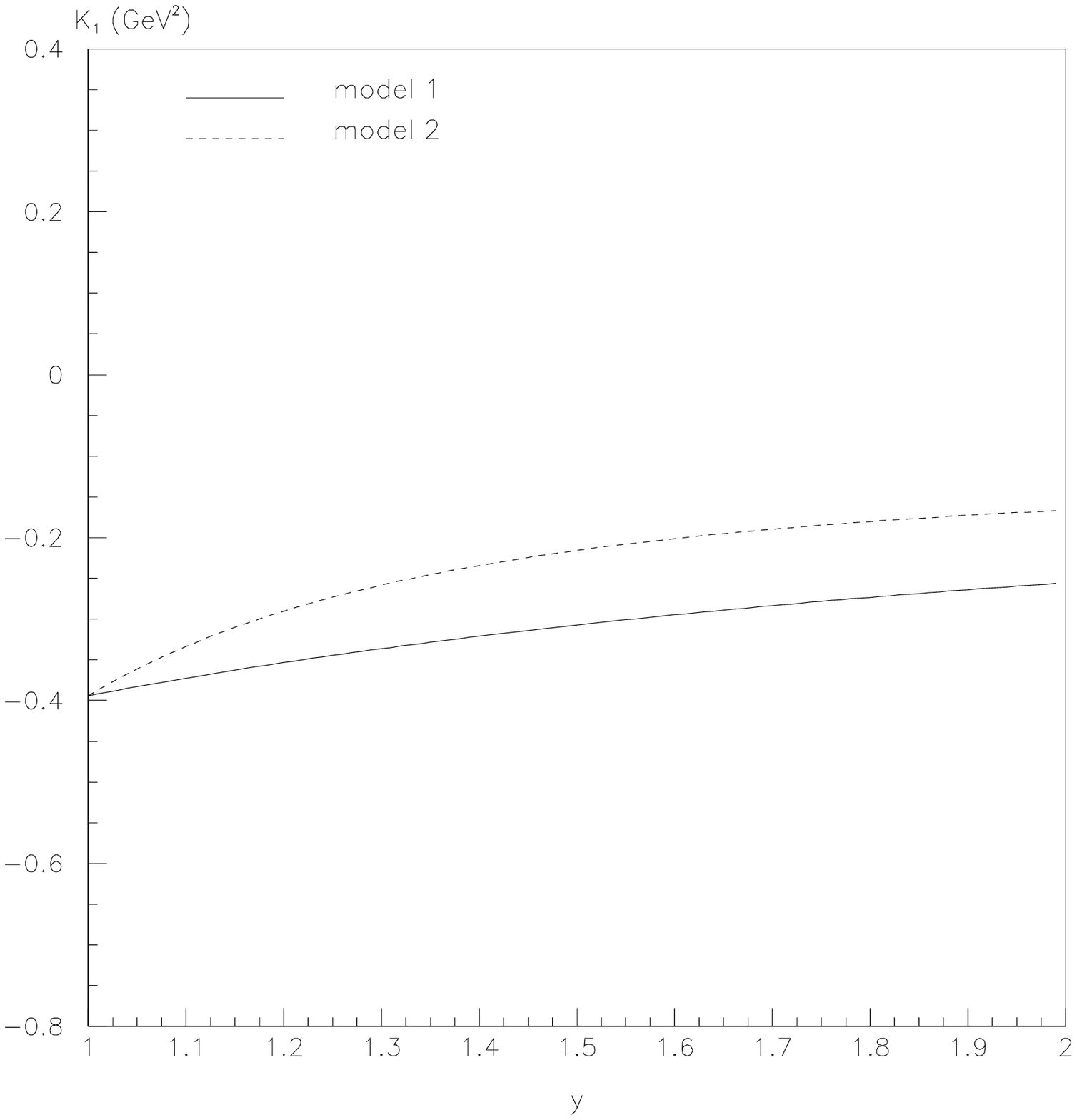}{ Fig.11. Sum rule result for $K_1$. }

\PIC{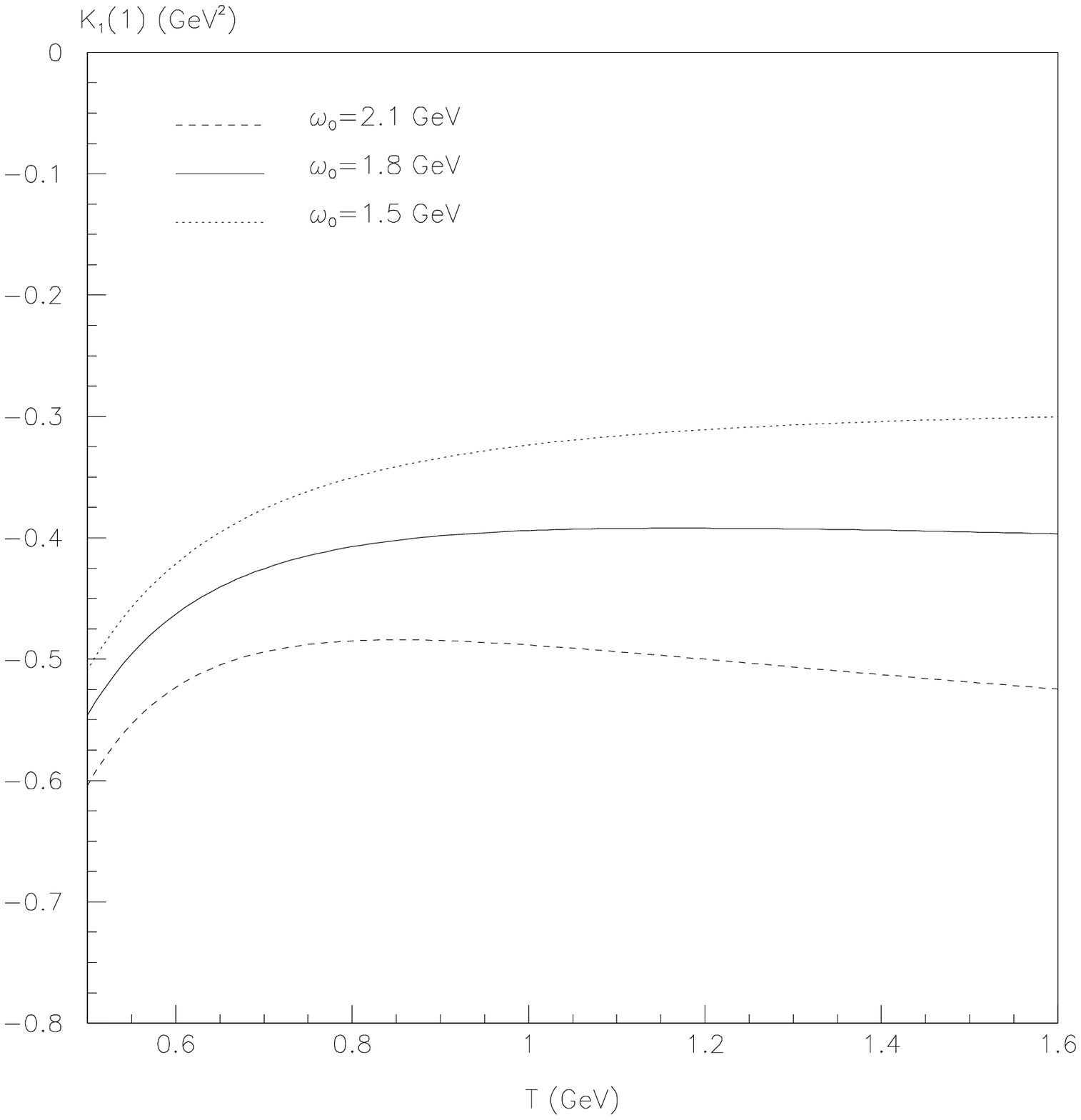}{ Fig.12. $K_1(1)$ as a function of Borel parameter $T$. }


\begin{center}
\begin{picture}(300,200)(0,0)

\SetWidth{2}
\Line(100,50)(150,130)
\Line(150,130)(200,50)
\Photon(150,160)(150,130){3}{2}
\SetWidth{0.5}
\Line(100,50)(200,50)
\Gluon(150,50)(150,130){3}{5}
\BBoxc(150,130)(8,8)
\DashLine(80,50)(100,50){3}
\DashLine(200,50)(220,50){3}


\end{picture}
\end{center}

\centerline{Fig.13. The lowest order perturbative diagram contributing to $\kappa_2$.}



\PIC{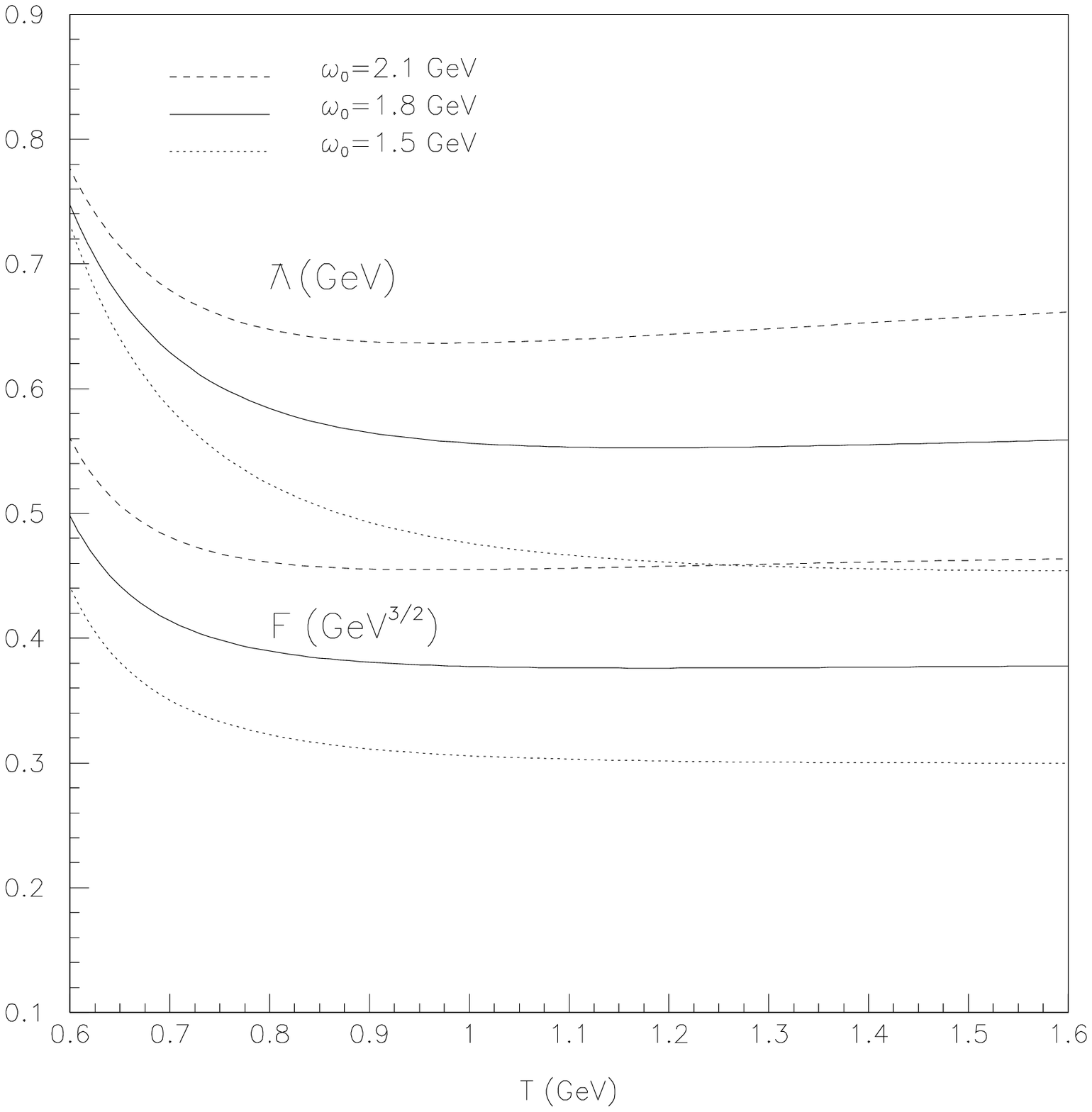}{ }

\vspace{-2.5cm}{\center{Fig.14. Leading oreder sum rule result for heavy meson decay when two-loop 
 perturbative contributions are considered.}
}

\vspace{2.5cm}

\PIC{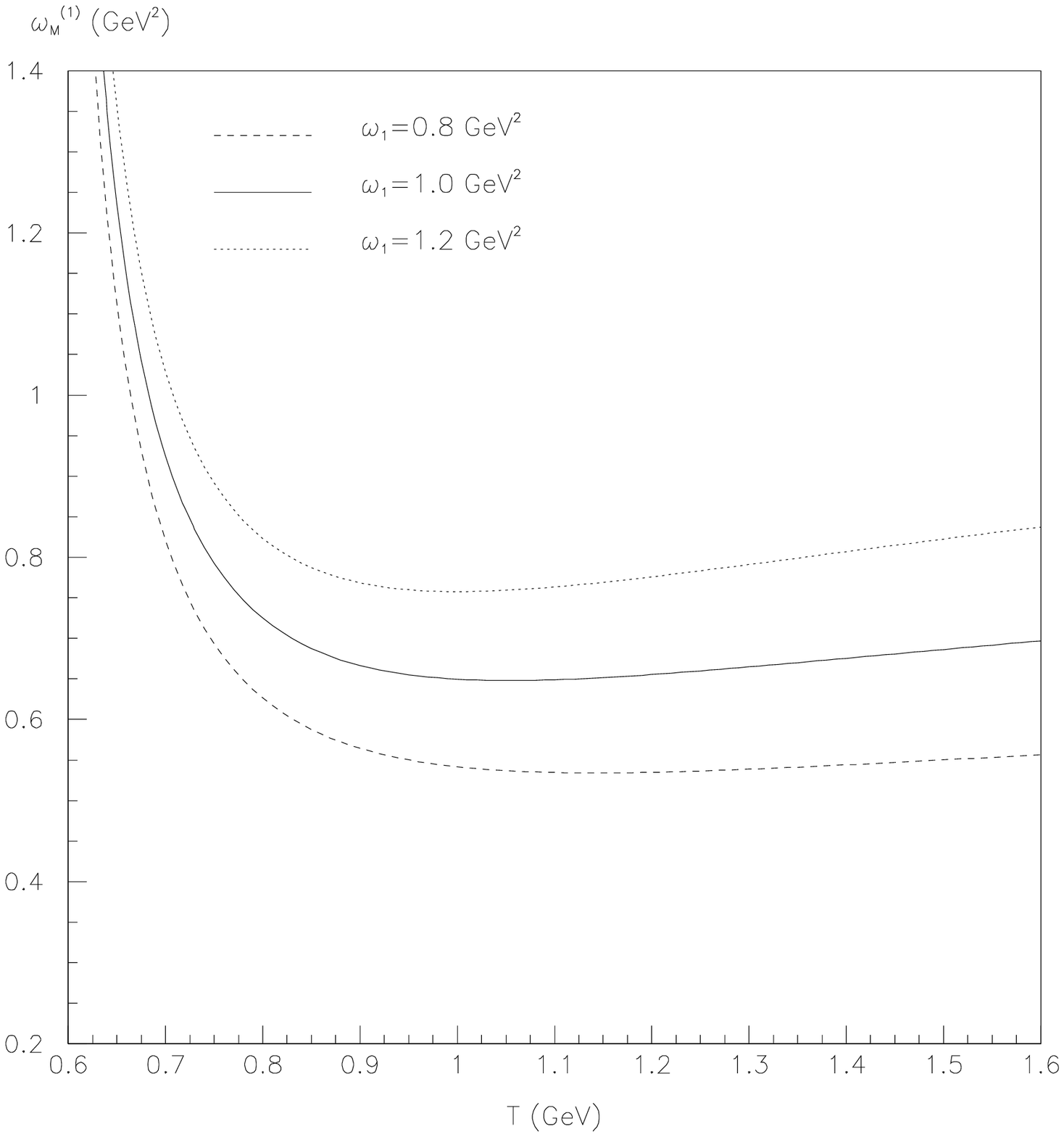}{Fig.15a}

\PIC{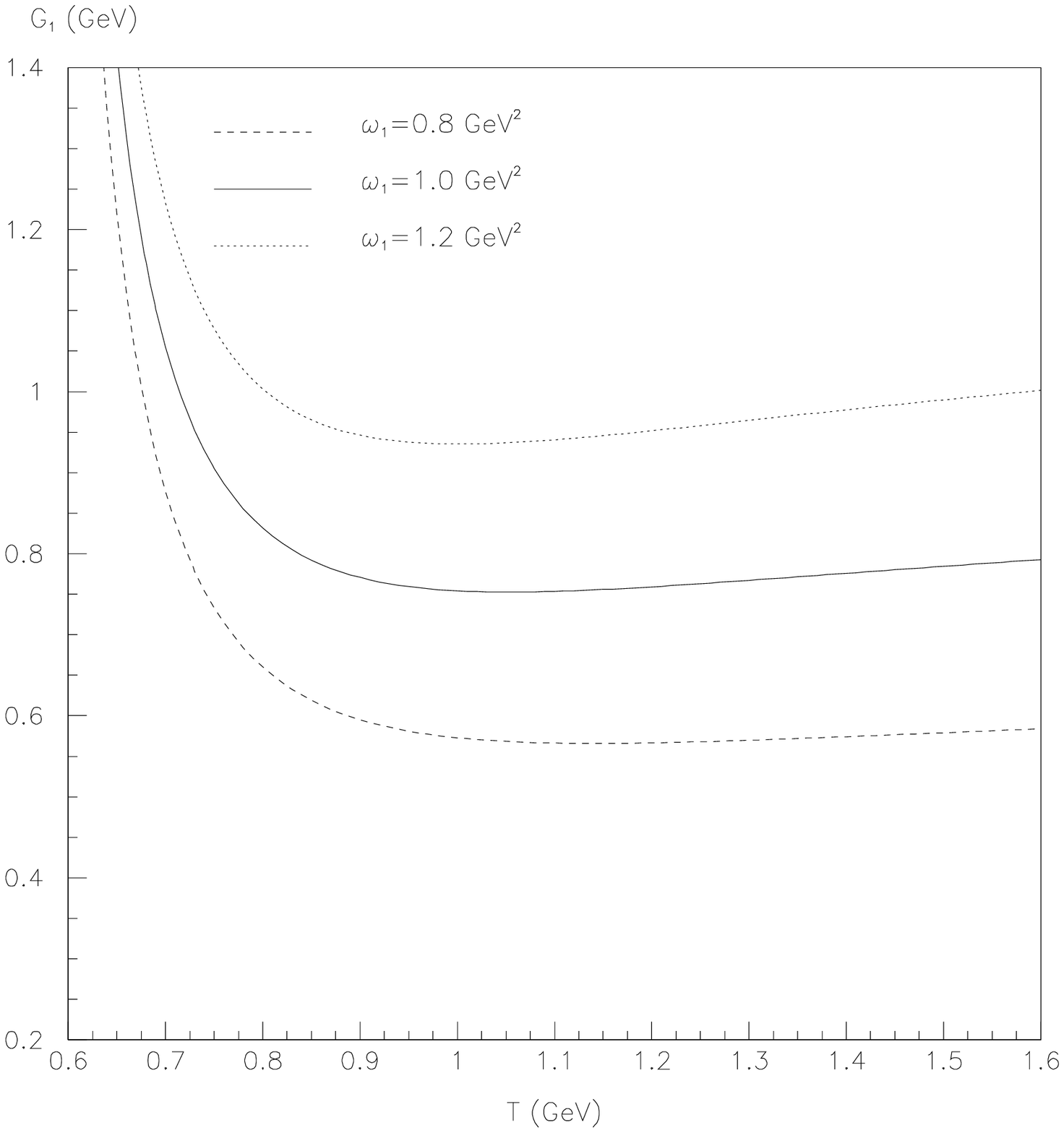}{Fig.15b}

\PIC{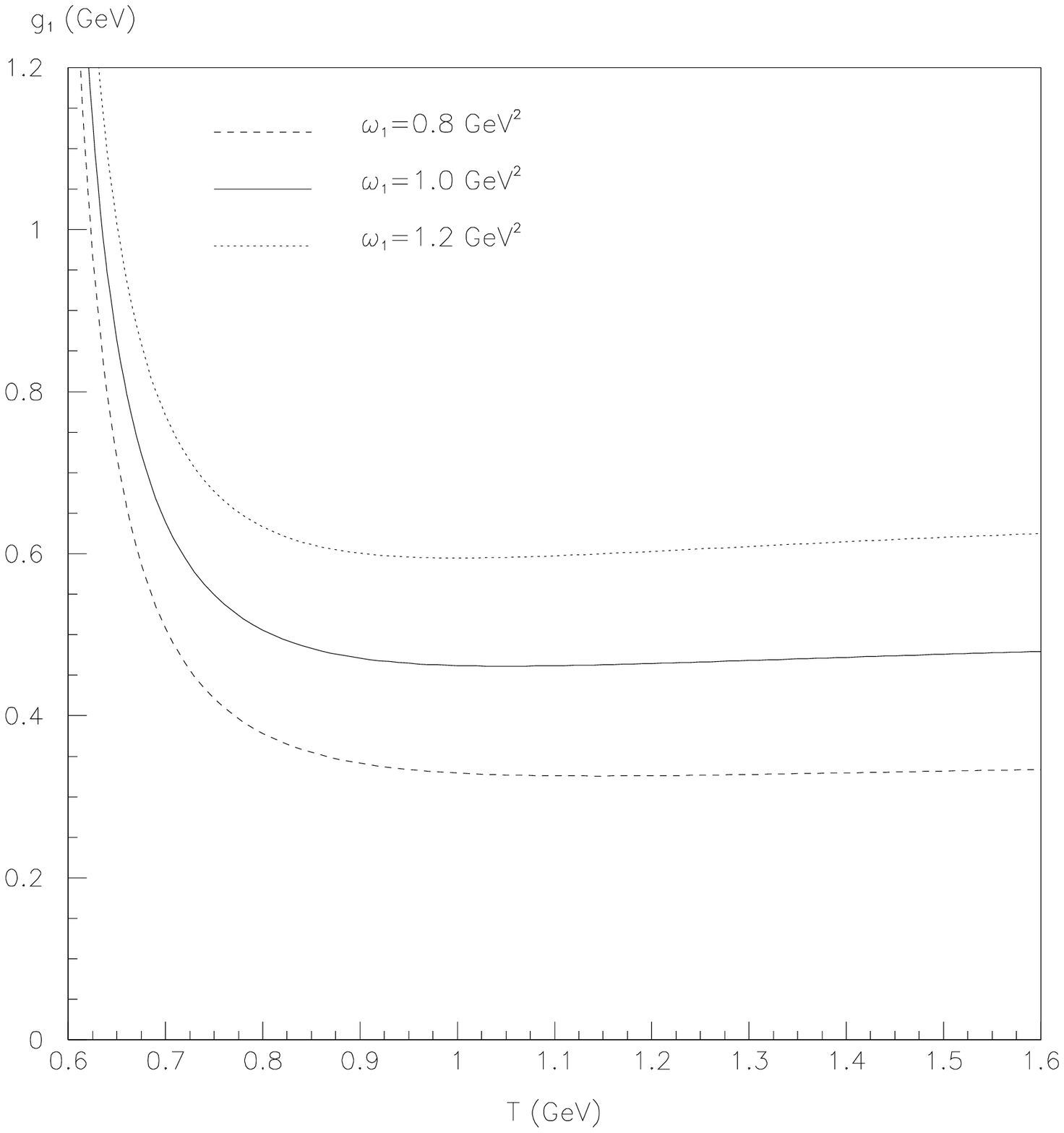}{Fig.15c}  

\vspace{-2cm}{\center{Fig.5 Sum rule result for $1/m_Q$ order spin-symmetry 
conserving corrections to heavy meson decay when two-loop perturbative contributions are 
considered.}
}

\vspace{2cm}

\PIC{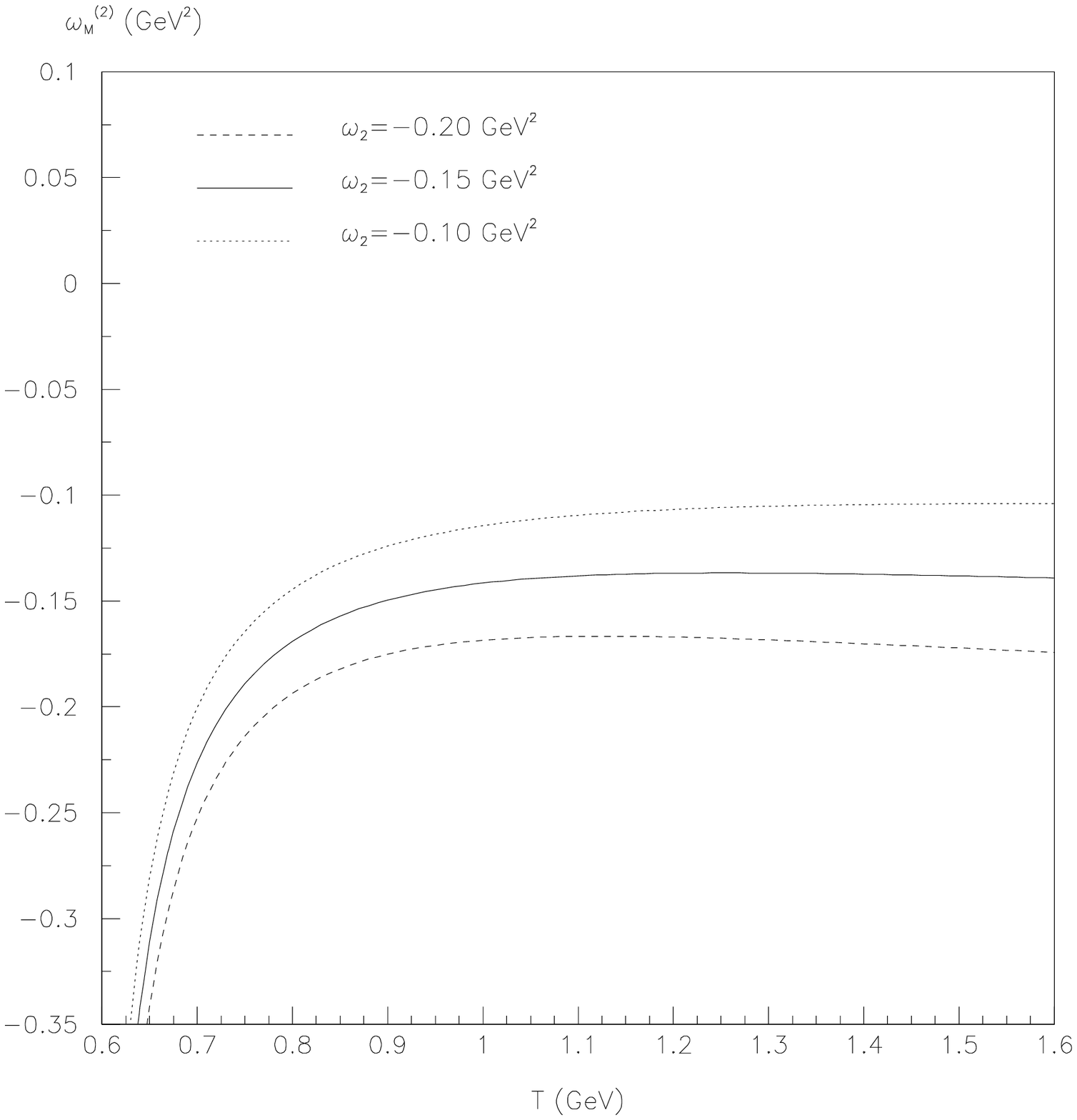}{Fig.16a}

\PIC{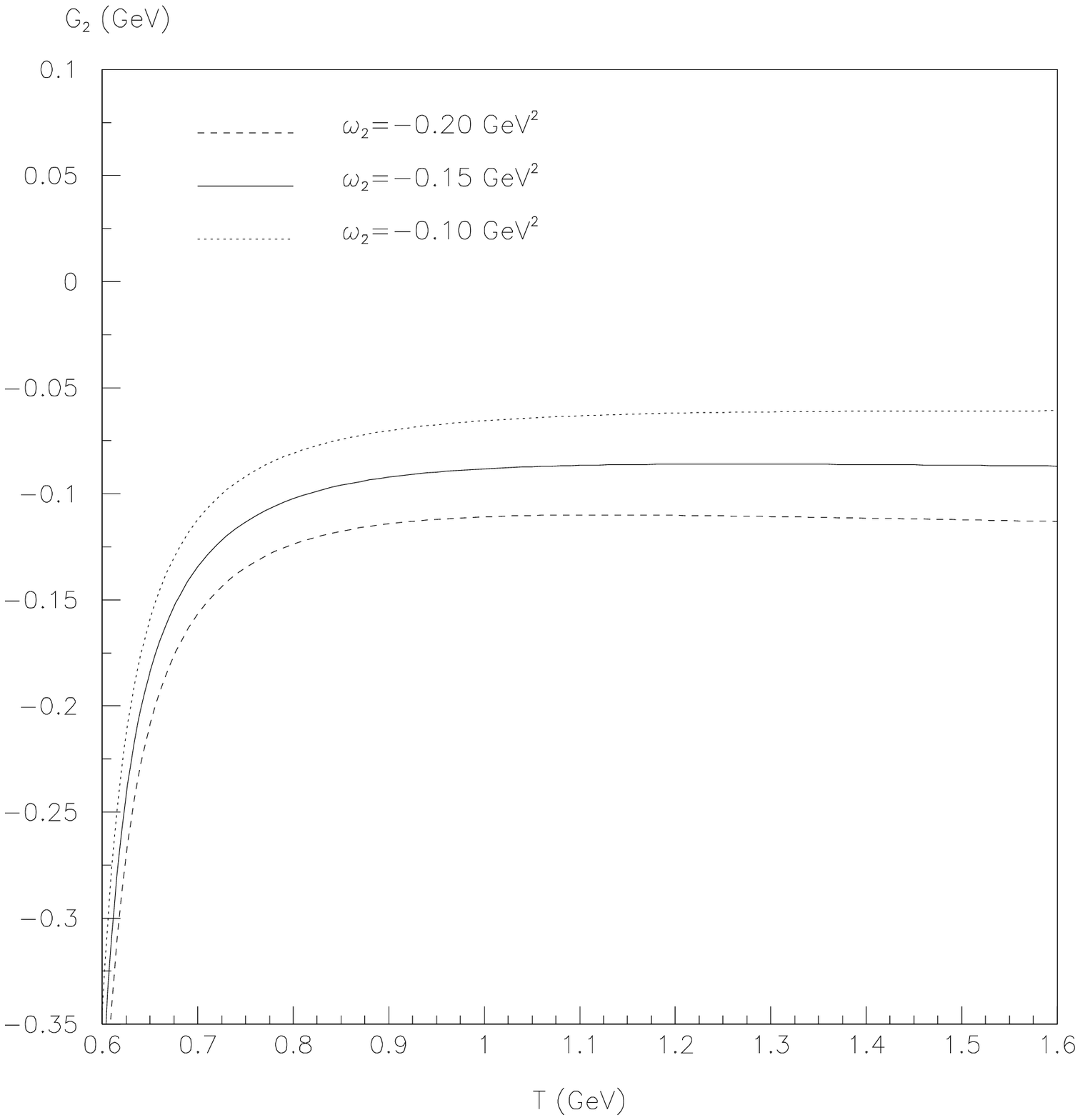}{Fig.16b}

\PIC{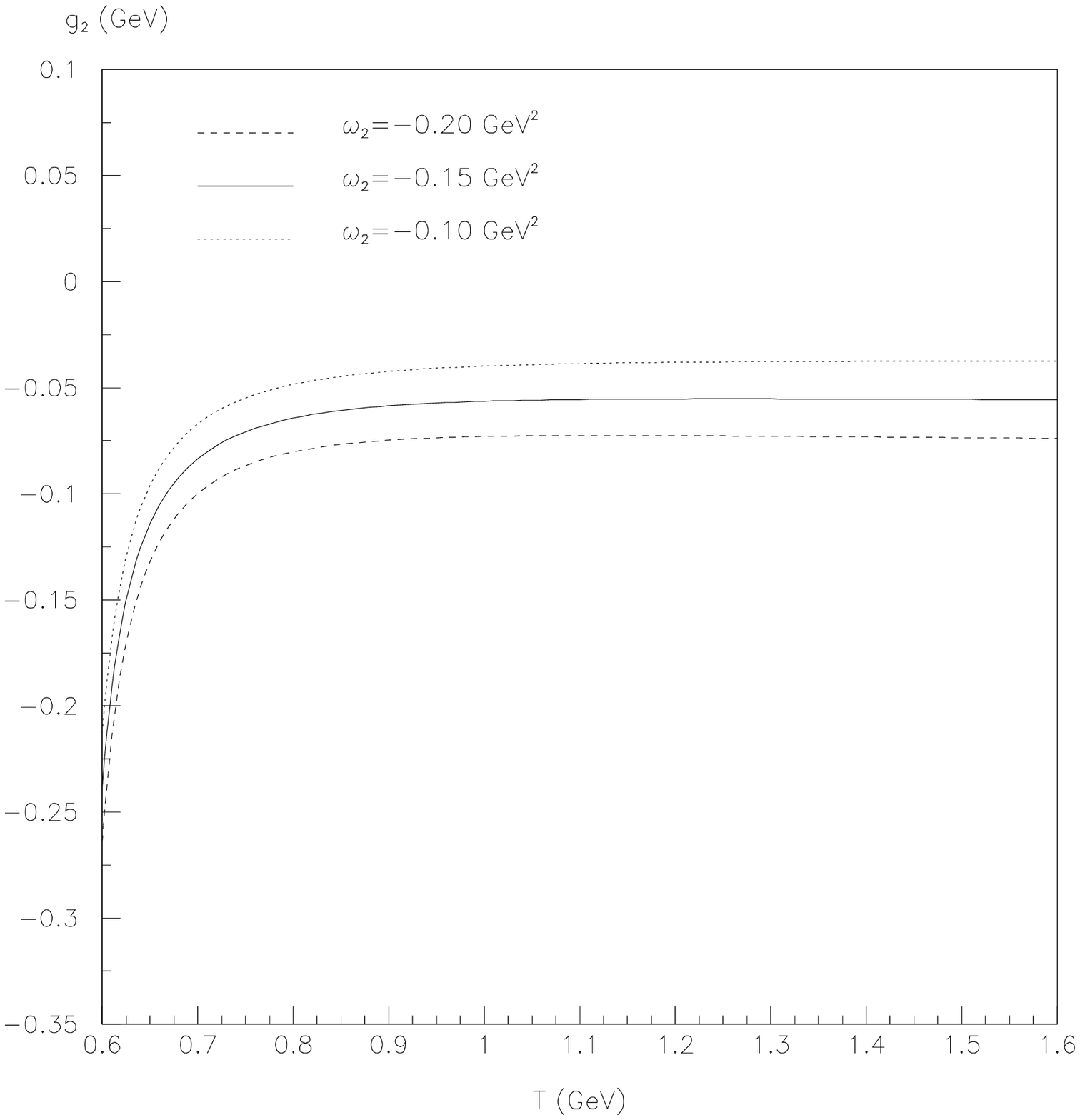}{Fig.16c}

\vspace{-1.95cm}{\center{Fig.6. Sum rule result for $1/m_Q$ order spin-symmetry 
breaking corrections to heavy meson decay when two-loop perturbative contributions are 
considered.} }

\vspace{0.77cm}

\PIC{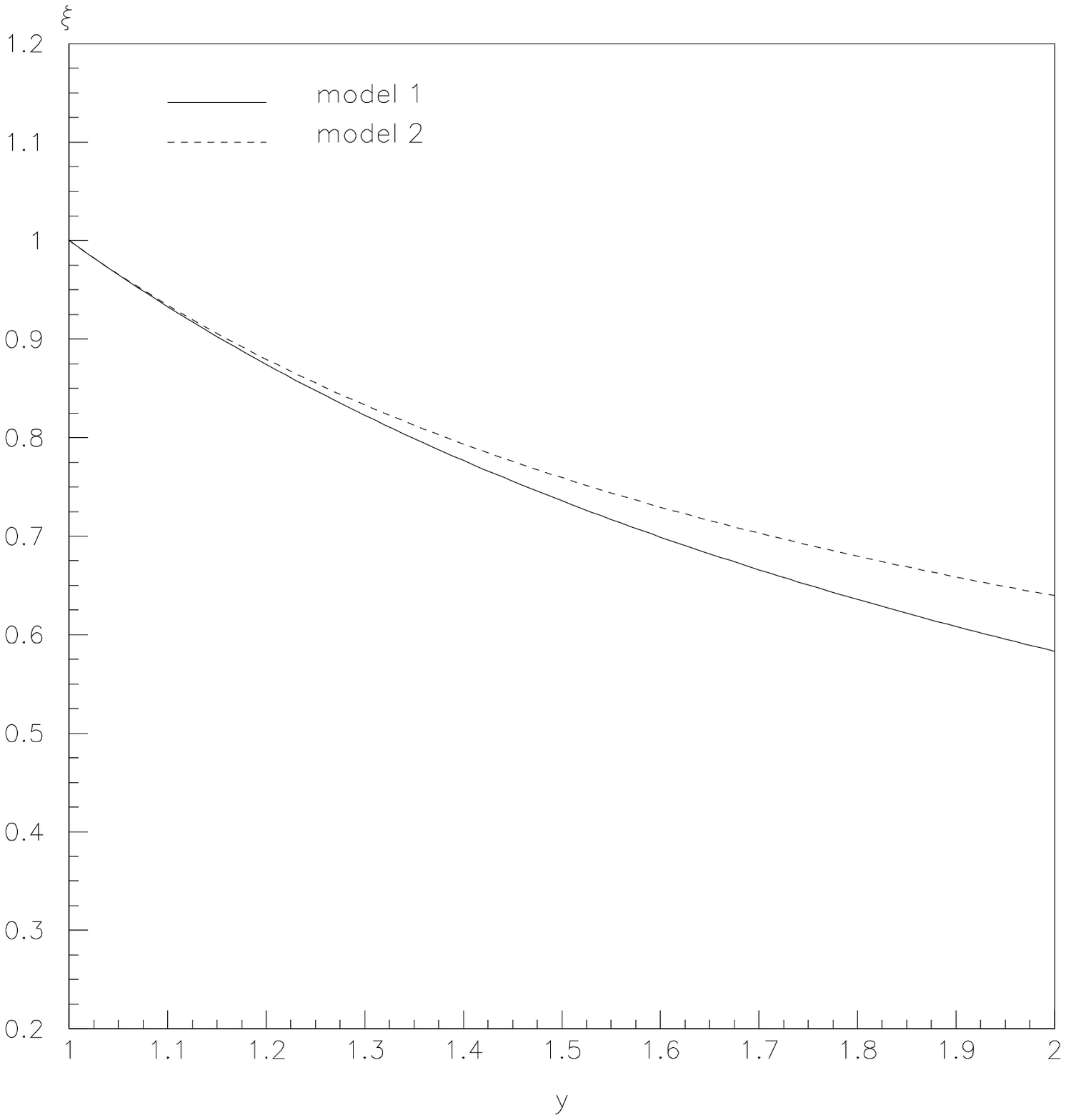}{ }

\vspace{-2.5cm}{\center{Fig.17. Sum rule result for Isgur-Wise function $\xi(y)$ when the 
 value of $F$ in eq.(\ref{eq:decval12loop}) is used as an input parameter.}
}

\vspace{2.5cm}

\PIC{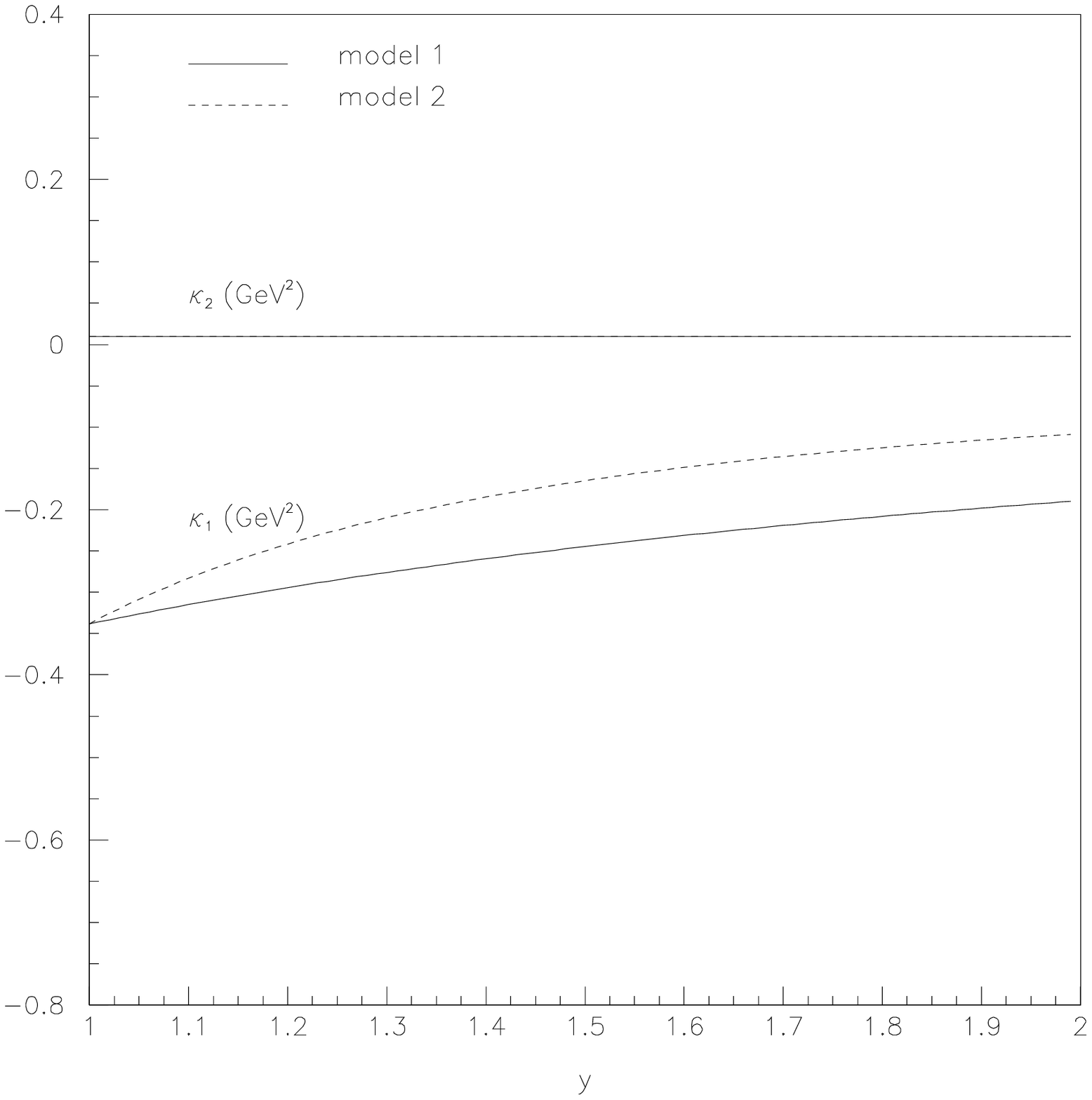}{ }

\vspace{-2.5cm}{\center{Fig.18. Sum rule result for $1/m_Q$ order transition form 
factor $\kappa_1$ and $\kappa_2$ when the value of $F$ in eq.(\ref{eq:decval12loop}) 
is used as an input parameter.}
}

\vspace{1.0cm}

\PIC{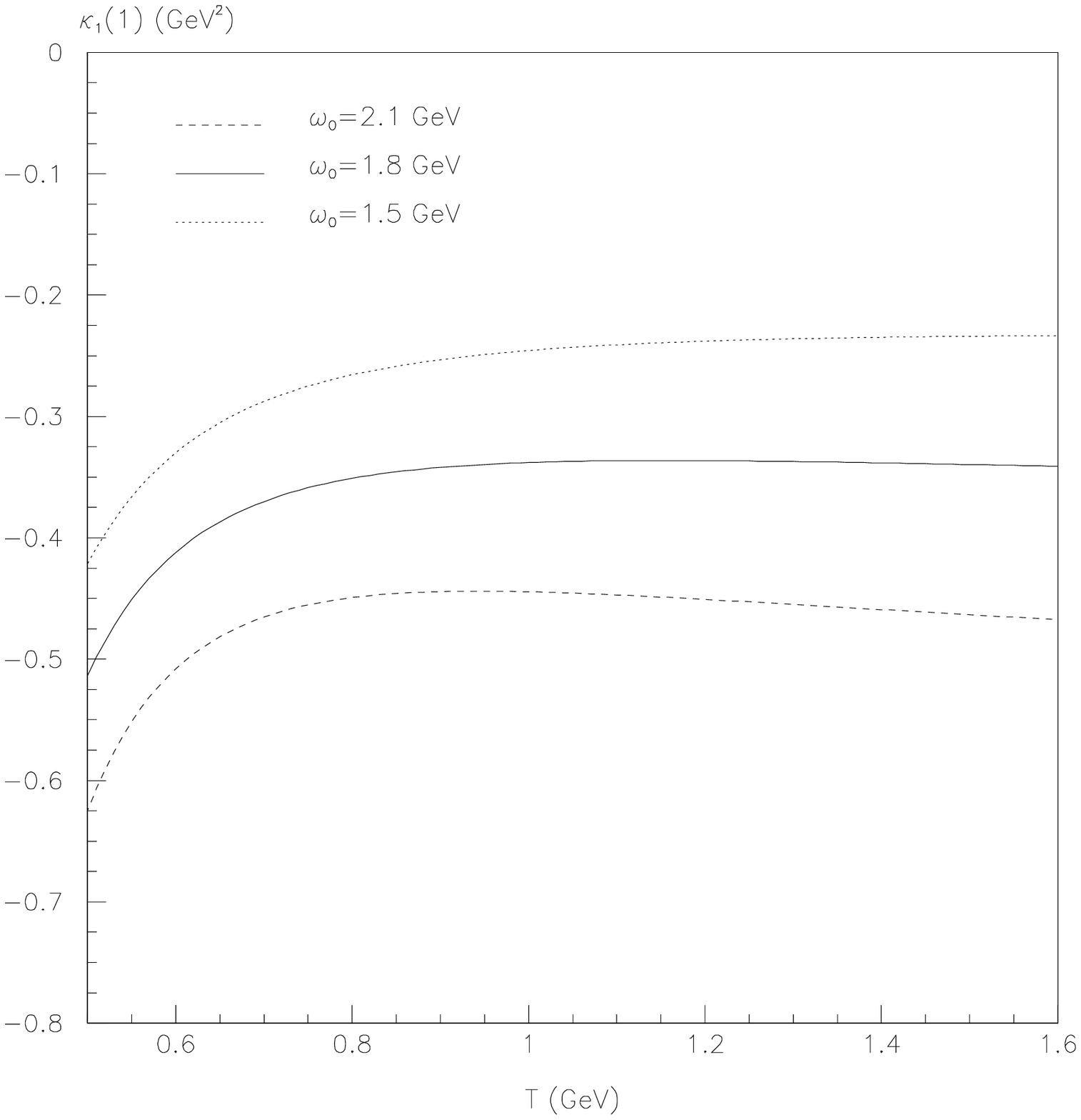}{ }

\vspace{-2.5cm}{\center{Fig.19. $\kappa_1(1)$ as a function of Borel parameter $T$ when the 
value of $F$ in eq.(\ref{eq:decval12loop}) is used as an input parameter.}
}

\vspace{2.5cm}

\PIC{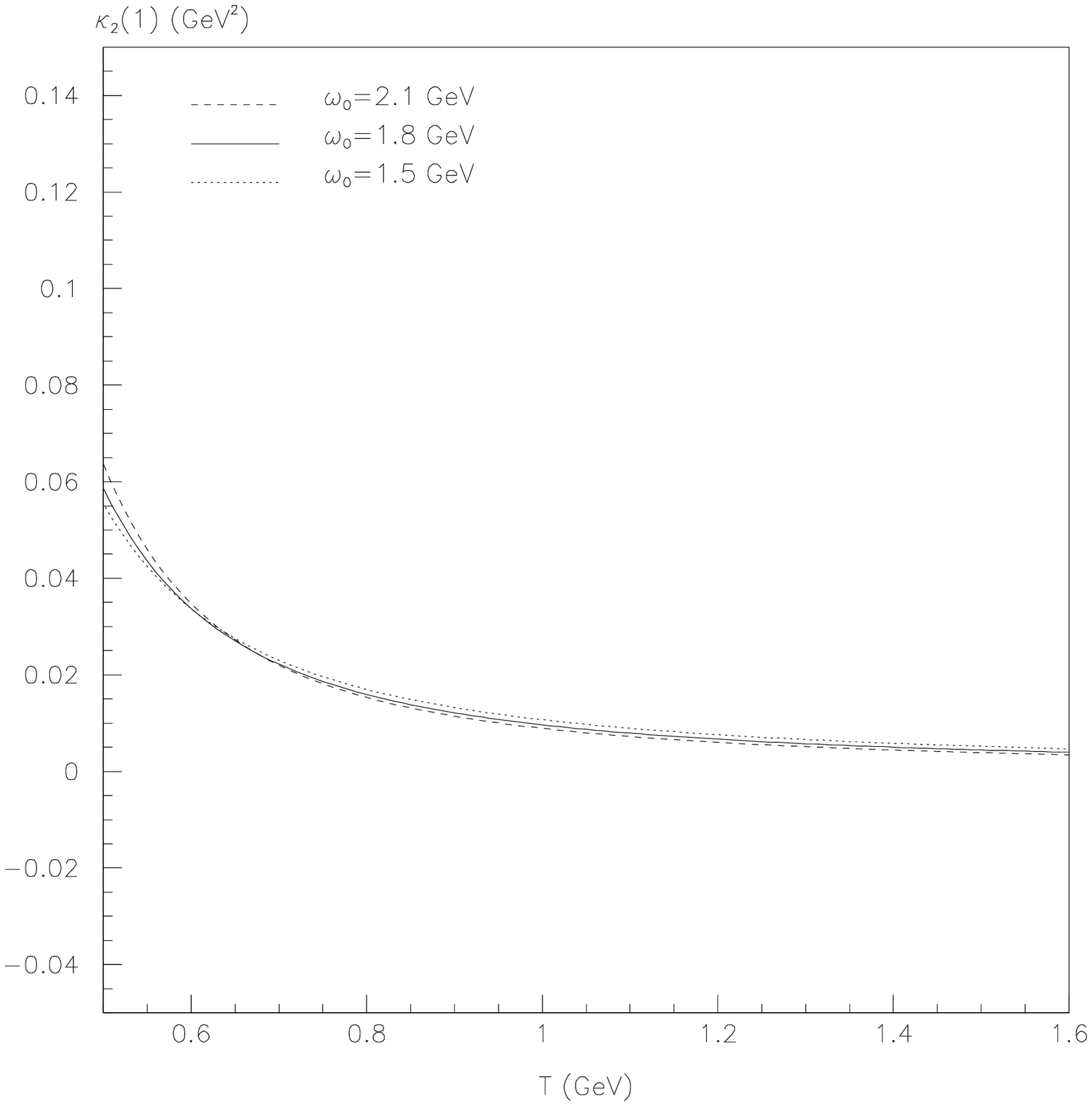}{ }

\vspace{-2.5cm}{\center{Fig.20. $\kappa_2(1)$ as a function of Borel parameter $T$ when the 
value of $F$ in eq.(\ref{eq:decval12loop}) is used as an input parameter.} }

\vspace{1.0cm}

\PIC{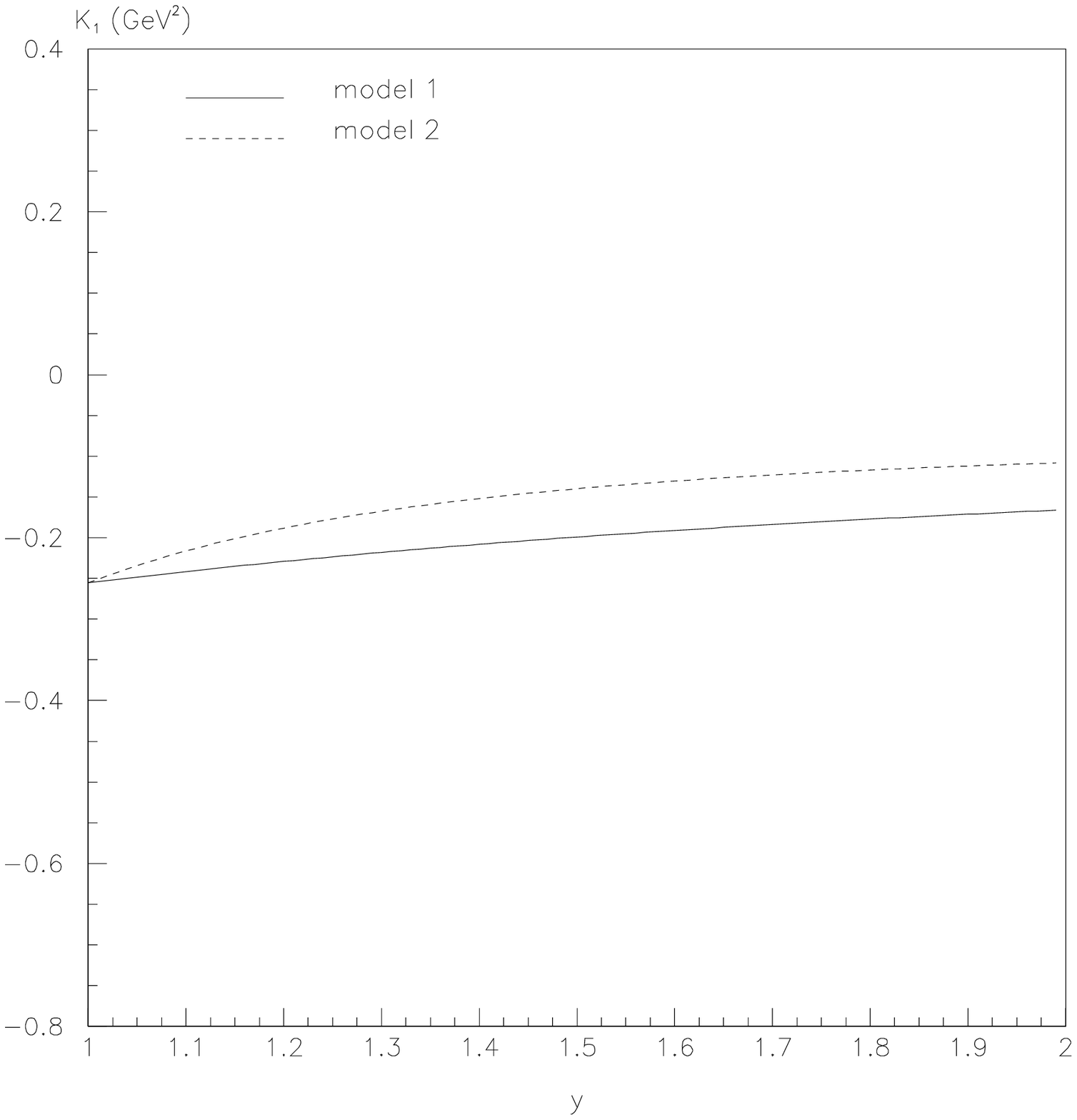}{ }

\vspace{-2.5cm}{\center{Fig.21. Sum rule result for $K_1$ when the 
value of $F$ in eq.(\ref{eq:decval12loop}) is used as an input parameter.}
}

\vspace{2.5cm}

\PIC{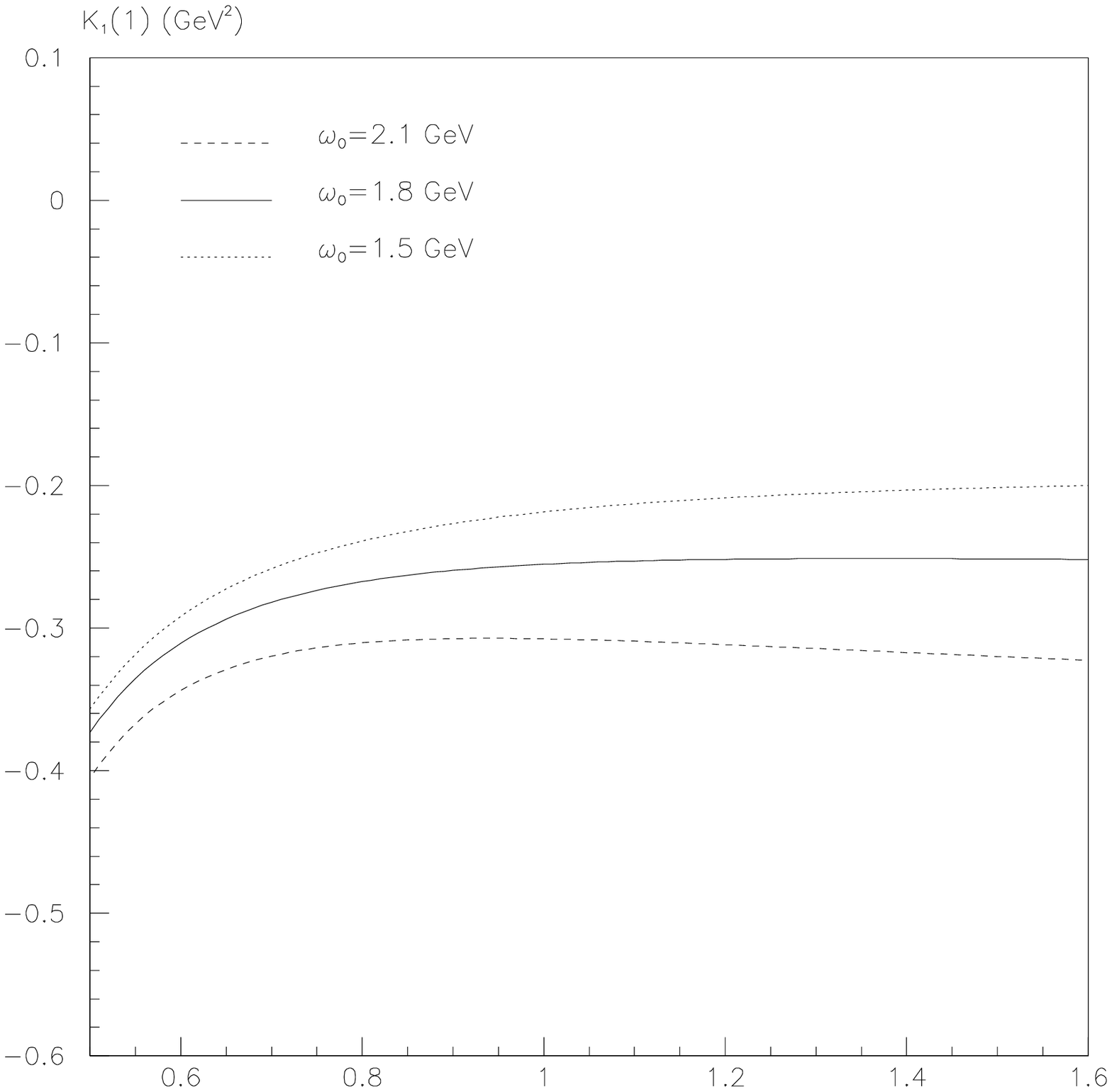}{ }

\vspace{-2.5cm}{\center{Fig.22. $K_1(1)$ as a function of Borel parameter $T$ when the 
value of $F$ in eq.(\ref{eq:decval12loop}) is used as an input parameter.}
}

\end{document}